\documentclass[journal]{IEEEtran}
\usepackage{ifpdf}
\usepackage{cite}
\usepackage[dvips]{graphicx}
\usepackage[cmex10]{amsmath}
\usepackage{amssymb}
\usepackage{amsmath}
\usepackage{mathrsfs}
\usepackage[lined,boxed,commentsnumbered, ruled]{algorithm2e}
\usepackage{array}
\usepackage{mdwmath}
\usepackage{eqparbox}
\usepackage[tight,footnotesize]{subfigure}

\newtheorem{proposition}{Proposition}
\newtheorem{definition}{Definition}

\usepackage[caption=false,font=footnotesize]{subfig}

\begin{document}

\title{Group Sparse Beamforming for Green Cloud-RAN}
\author{Yuanming~Shi,~\IEEEmembership{Student Member,~IEEE,}
        Jun~Zhang,~\IEEEmembership{Member,~IEEE,}
        and~Khaled~B. Letaief,~\IEEEmembership{Fellow,~IEEE}
\thanks{The authors are with the Department of Electronic and Computer Engineering, the Hong Kong University of Science and Technology, Clear Water Bay, Kowloon, Hong Kong. E-mail: $\{\textrm{yshiac, eejzhang, eekhaled}\}$@ust.hk.}}


\maketitle

\begin{abstract}
A cloud radio access network (Cloud-RAN) is a network architecture that holds the promise of meeting the explosive growth of mobile data traffic. In this architecture, all the baseband signal processing is shifted to a single baseband unit (BBU) pool, which enables efficient resource allocation and interference management. Meanwhile, conventional powerful base stations can be replaced by low-cost low-power remote radio heads (RRHs), producing a green and low-cost infrastructure. However, as all the RRHs need to be connected to the BBU pool through optical transport links, the transport network power consumption becomes significant. In this paper, we propose a new framework to design a green Cloud-RAN, which is formulated as a joint RRH selection and power minimization beamforming problem. To efficiently solve this problem, we first propose a greedy selection algorithm, which is shown to provide near-optimal performance. To further reduce the complexity, a novel group sparse beamforming method is proposed by inducing the group-sparsity of beamformers using the weighted $\ell_1/\ell_2$-norm minimization, where the group sparsity pattern indicates those RRHs that can be switched off. Simulation results will show that the proposed algorithms significantly reduce the network power consumption and demonstrate the importance of considering the transport link power consumption.

 \end{abstract}

\begin{IEEEkeywords}
Cloud-RAN, green communication, power consumption, greedy
selection, group-sparsity. 
\end{IEEEkeywords}

\IEEEpeerreviewmaketitle

\section{Introduction}
\IEEEPARstart{M}{obile} data traffic has been growing enormously in recent years, and it is expected that cellular networks will have to offer a 1000x increase in capacity in the following decade to meet this demand \cite{Soliman_CM13}. Massive MIMO \cite{Rusek_SPM2013} and heterogeneous and small cell networks (HetSNets) \cite{Soliman_CM13} are regarded as two most promising approaches to achieve this goal. By deploying a large number of antennas at each base station (BS), massive MIMO can exploit spatial multiplexing gain in a large-scale and also improve energy efficiency. However, the performance of massive MIMO is limited by correlated scattering with the antenna spacing constraints, which also brings high deployment cost to maintain the minimum spacing \cite{Soliman_CM13}. HetSNets exploit the spatial reuse by deploying more and more access points (APs). Meanwhile, as stated in \cite{Debbah_VTM11}, placing APs based on the traffic demand is an effective way for compensating path-loss, resulting in energy efficient cellular networks. However, efficient interference management is challenging for dense small-cell networks. Moreover, deploying more and more small-cells will cause significant cost and operating challenges for operators.  

Cloud radio access network (Cloud-RAN) has recently been proposed as a promising network architecture to unify the above two technologies in order to jointly manage the interference (via coordinated multiple-point process
(CoMP)), increase network capacity and energy efficiency (via network densification),
and reduce both the network capital expenditure (CAPEX) and operating expense
(OPEX) (by moving baseband processing to the baseband unit (BBU) pool) \cite{mobile2011c,Wu_WC2012}. A large-scale distributed cooperative MIMO system will thus be formed.  Cloud-RAN can therefore be regarded as the ultimate solution to the ``spectrum crunch" problem of cellular networks.

There are three key components in a Cloud-RAN: (i) a pool of BBUs in a datacenter \emph{cloud}, supported by
the real-time virtualization and high performance processors, where all the baseband processing is performed; (ii) a high-bandwidth low-latency optical transport network connecting the BBU pool and the remote radio heads (RRHs); and (iii) distributed transmission/reception points (i.e., RRHs).  The key feature of Cloud-RAN is that RRHs and BBUs are separated, resulting a centralized BBU pool, which enables efficient cooperation of the transmission/reception among different RRHs. As a result, significant performance improvements through joint scheduling and joint signal processing such as coordinated beamforming or multi-cell processing\cite{Gesbert_JSAC10} can be achieved. With efficient interference suppression, a network of RRHs with a very high density can be deployed. This will also reduce the communication distance to the mobile terminals and can thus significantly reduce the transmission power. Moreover, as baseband signal processing is shifted to the BBU pool, RRHs only need to support basic transmission/reception functionality, which further reduces their energy consumption and deployment cost. 

The new architecture of Cloud-RAN also indicates a paradigm shift in the network design, which causes some technical challenges for implementation. For instance, as the data transmitted between the RRHs and the BBU pool is typically oversampled real-time I/Q digital data streams in the order of Gbps, high-bandwidth optical transport links with low-latency will be needed. To support CoMP and computing resource sharing among BBUs, new virtualization technologies need to be developed to distribute or group the BBUs into a centralized entity \cite{mobile2011c}. Another important aspect is the energy efficiency consideration, due to the increased power consumption of a large number of RRHs and also of the transport links. 

Conventionally, the transport network (i.e., backhaul links between the core network and base stations (BSs)) power consumption can be ignored as it is negligible compared to the power consumption of macro  BSs. Therefore, all the previous works investigating the energy efficiency of cellular networks only consider the BS power consumption \cite{WeiYu_WC10,Chang_ICC2013}. Recently, the impact of the backhaul power consumption in cellular network was investigated in \cite{Tombaz_GLOBECOM2011}, where it was shown through simulation that the backhaul power consumption will affect the energy efficiency of different cellular network deployment scenarios. Subsequently, Rao \emph{et al} in \cite{Rao_VT2013} investigated the spectral efficiency and energy efficiency tradeoff in homogeneous cellular
networks when taking the backhaul power consumption into consideration. 

In
Cloud-RAN, the transport network power consumption will have a more
significant impact on the network energy efficiency. Hence, allowing
the transport links and the corresponding RRHs to support the sleep mode will be
essential
to reduce the network power consumption for the Cloud-RAN. Moreover, with
the spatial and temporal variation of the mobile traffic, it would be feasible
to switch off some RRHs while still maintaining the QoS requirements. It will be also practical to implement such idea in the Cloud-RAN with the help of  centralized signal processing at the BBU pool. As energy efficiency is one of the major objectives for future cellular networks \cite{Wu_WC2012}, in this paper we will focus on the design of green Cloud-RAN by jointly considering
the power consumption of the transport network and RRHs.

\subsection{Contributions}
The main objective of this paper is to minimize the network power consumption of Cloud-RAN, including the transport network and radio access network power consumption, with a quality of service (QoS) constraint at each user. Specifically, we formulate the design problem as a joint RRH selection and power minimization beamforming problem, where the transport network power consumption is determined by the set of active RRHs, while the transmit power consumption of the active RRHs is minimized through coordinated beamforming. This is a mixed-integer non-linear programming (MINLP) problem, which is NP-hard. We will focus on designing low-complexity algorithms for practical implementation. The major contributions of the paper are summarized as follows:
\begin{enumerate}
\item We formulate the network power consumption minimization problem for the Cloud-RAN by enabling both the transport links and RRHs to support the sleep mode.  In particular, we provide a group sparse beamforming (GSBF) formulation of the design problem, which assists the problem analysis and algorithm design. 

\item We first propose a greedy selection  (GS) algorithm, which selects one RRH to switch off at each step. It turns out that the RRH selection rule is critical, and we propose to switch off the RRH that \emph{maximizes the reduction in the network power consumption} at each step. From the simulations, the proposed GS algorithm often yields optimal or near-optimal solutions, but its complexity may still be prohibitive for a large-size  network.  

\item To further reduce the complexity, we propose a three-stage group sparse beamforming (GSBF) framework, by adopting the weighted mixed $\ell_{1}/\ell_{p}$-norm to induce the group sparsity for
the beamformers. In contrast to all the previous works applying the mixed $\ell_{1}/\ell_{p}$-norm
to induce group sparsity, we exploit the additional prior information (i.e., transport power consumption, power amplifier efficiency, and instantaneous effective channel gain) to design the weights for different beamformer coefficient groups, resulting in a significant performance gain. Two GSBF algorithms with different complexities are proposed: namely, a bi-section GSBF algorithm and an iterative GSBF algorithm.

\item We shall show that the GS algorithm always provides near-optimal performance. Hence, it would be a good option if the number of RRHs is relatively small, such as in clustered deployment. With a very low computational complexity, the bi-section GSBF algorithm is an attractive option for a large-scale
Cloud-RAN. The iterative GSBF algorithm provides a good tradeoff between complexity and performance, which makes it a good candidate for a medium-size network. 

\end{enumerate}

\subsection{Related Works}

A main design tool applied in this paper is optimization with the group sparsity induced norm. With the recent theoretical breakthrough in compressed sensing \cite{Donoho_TIT2006, Tao_IT06}, the sparsity patterns in different applications in signal processing and communications have been exploited for more efficient system design, e.g., for pilot aided sparse channel estimation \cite{Berger_CM10}. 
The sparsity inducing norms have been widely applied
in high-dimensional statistics, signal processing, and machine learning in
the last decade \cite{Bach_ML2011}. The $\ell_{1}$-norm regularization has
been successfully applied in compressed sensing \cite{Donoho_TIT2006, Tao_IT06}. More recently, mixed $\ell_{1}/\ell_{p}$-norms
are widely investigated in the case where some variables forming a group
will be selected or removed simultaneously, where the mixed $\ell_{1}/\ell_{2}$-norm
\cite{Ming_SM06} and mixed $\ell_{1}/\ell_{\infty}$-norm \cite{Wainwright_IT2011}
are two commonly used ones to induce group sparsity for their computational and
analytical convenience. 

In Cloud-RAN, one RRH will be switched off only when all the coefficients in its beamformer are set to zeros.  In other words, all the coefficients in the beamformer at one RRH should be selected or ignored simultaneously, which requires group sparsity rather than individual sparsity for the coefficients as commonly used in compressed sensing. In this paper, we will adopt the mixed $\ell_{1}/\ell_{p}$-norm to promote group sparsity for the beamformers instead of $\ell_1$-norm, which only promotes individual sparsity. Recently, there are some works \cite{Z.Q.Luo_JSAC2013, TonyQ.S._WC2013, Mehanna_SP2013} adopting the mixed $\ell_{1}/\ell_{p}$-norm to induce group-sparsity in a large-scale cooperative wireless cellular network. Specifically, Hong {\emph{et al.}} \cite{Z.Q.Luo_JSAC2013} adopted the mixed $\ell_{1}/\ell_{2}$-norm and Zhao \emph{et al.}  \cite{TonyQ.S._WC2013} used the $\ell_{2}$-norm to induce the group sparsity of the beamformers, which reduce the amount of the shared user data among different BSs. The squared mixed $\ell_{1}/\ell_{\infty}$-norm was investigated in \cite{Mehanna_SP2013} for antenna selection. 

All of the above works simply adopted the un-weighted mixed $\ell_{1}/\ell_{p}$-norms to induce group-sparsity, in which, no prior information on the unknown signal is assumed other than the fact that it is
sufficiently sparse. By exploiting the prior information in terms of system parameters, the weights for different beamformer coefficient
groups can be more rigorously designed and performance can be enhanced. We demonstrate through simulations that
the proposed three-stage GSBF framework, which is based on the weighted mixed $\ell_{1}/\ell_{p}$-norm minimization, outperforms the conventional unweighted
mixed $\ell_{1}/\ell_{p}$-norm minimization based algorithms substantially.

\subsection{Organization}
The remainder of the paper is organized as follows. Section II presents the system and power model. In section III, the network power consumption minimization problem is formulated, followed by some analysis. Section IV presents the GS algorithm, which yields near-optimal solutions. The three-stage GSBF framework is presented in Section V. Simulation results will be presented in Section VI. Finally, conclusions and discussions are presented in Section VII. 

\emph{Notations}: $\|\cdot\|_{\ell_{p}}$ is the $\ell_{p}$-norm. Boldface lower case and upper case letters represent vectors and matrices, respectively.  $(\cdot)^{T}$, $(\cdot)^{\dag}$, $(\cdot)^{\sf{H}}$ and $\textrm{Tr}(\cdot)$ denote the transpose, conjugate,  Hermitian and trace operators, respectively. $\mathfrak{R}(\cdot)$ denotes the real part. 
 
\section{System and Power Model}
\subsection{System Model}
\begin{figure}[t]
 \centering
 \includegraphics[width=1\columnwidth]{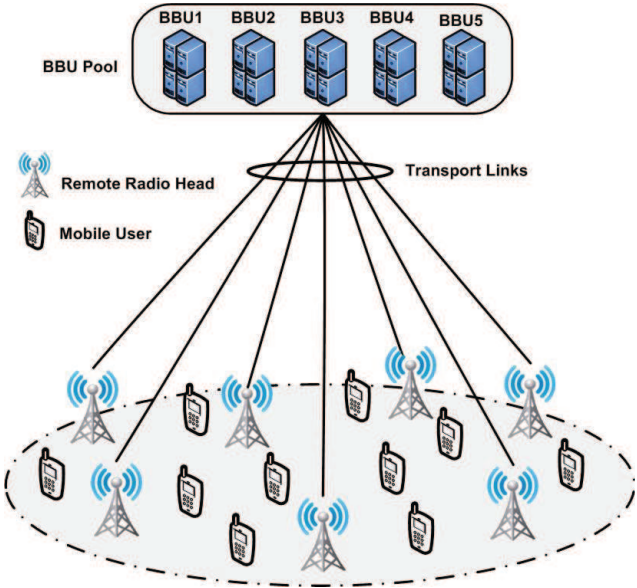}
\caption{The architecture of Cloud-RAN, in which, all the RRHs are connected
to a BBU pool through transport links.}
\label{system}
 \end{figure}

We consider a Cloud-RAN with $L$ remote radio heads (RRHs),  where the $l$-th RRH is equipped with $N_{l}$ antennas, and $K$ single-antenna mobile users (MUs), as shown in Fig. {\ref{system}}.
In this network architecture, all the base band units (BBUs) are moved into a single BBU pool, creating a set of shared processing resources, and enabling efficient interference management and mobility management. With the baseband signal processing functionality migrated to the BBU pool, the RRHs can be deployed in a large scale with low-cost. The BBU pool is connected to the RRHs using the common public radio interface (CPRI) transport technology via  a high-bandwidth, low-latency optical transport network \cite{mobile2011c}. The digitized baseband complex inphase (I) and quadrature (Q) samples of the radio signals are transported over the transport links between the BBUs and RRHs. The key technical and economic issue of the Cloud-RAN is that this architecture requires significant transport network resources. As the focus of this paper is on network power consumption, we will assume all the transport links have sufficiently high capacity and negligible latency\footnote{The impact of limited-capacity transport links on compression in Cloud-RAN was recently investigated in \cite{Shamai_TVT2013,Shamai_TSP2013}, and its impact in our setting is left to future work.}.

Due to the high density of RRHs and the joint transmission among them, the energy used for signal transmission will be reduced significantly. However,
the power consumption of the transport network becomes enormous and cannot be ignored. Therefore, it is highly desirable to switch off some transport links and the corresponding  RRHs to reduce the network power consumption based on the data traffic requirement, which forms the main theme of this work.   

Let $\mathcal{L}=\{1,...,L\}$ denote the set of RRH indices, $\mathcal{A}\subseteq\mathcal{L}$ denote the active RRH set, $\mathcal{Z}$ denote the inactive RRH set with $\mathcal{A}\cup\mathcal{Z}=\mathcal{L}$, and $\mathcal{S}=\{1,...,K\}$ denote the index set of scheduled users. In a beamforming design framework, the baseband transmit signals are of the form:
\begin{eqnarray}
{\bf{x}}_{l}=\sum\limits_{k=1}^{K}{\bf{w}}_{lk}s_{k}, \forall l\in\mathcal{A},
\end{eqnarray}
where $s_{k}$ is a complex scalar denoting the data symbol for user
$k$ and  ${\bf{w}}_{lk}\in\mathbb{C}^{N_{l}}$ is the beamforming vector at RRH $l$ for user $k$. Without loss of generality, we assume that $E[|s_{k}|^2]=1$ and $s_{k}$'s are independent with each other. The baseband signals ${\bf{x}}_l$'s will be transmitted to the corresponding RRHs, but not the data information $s_{k}$'s \cite{mobile2011c,Shamai_TSP2013}. The baseband received signal at user $k$ is given by
\begin{eqnarray}
y_{k}=\sum\limits_{l\in\mathcal{A}}{\bf{h}}_{kl}^{\sf{H}}{\bf{w}}_{lk}s_{k}+\sum\limits_{i\ne
k}\sum\limits_{l\in\mathcal{A}}{\bf{h}}_{kl}^{\sf{H}}{\bf{w}}_{li}s_{i}+z_{k}, k\in\mathcal{S},
\end{eqnarray}
where ${\bf{h}}_{kl}\in\mathbb{C}^{N_{l}}$ is the channel vector
from RRH $l$ to user $k$,  and $z_{k}\sim\mathcal{CN}(0,\sigma_{k}^2)$
is the additive Gaussian noise.

We assume that all the users are employing single user detection (i.e., treating interference as noise), so that they can use the receivers with low-complexity and energy-efficient structure. Moreover, in the low interference region, treating interference as noise can be optimal \cite{Jafar_IT2008}. The corresponding signal-to-interference-plus-noise ratio (SINR) for user $k$ is hence given by
\begin{eqnarray}
{\rm{SINR}}_{k}={{|\sum_{l\in\mathcal{A}}{\bf{h}}_{kl}^{\sf{H}}{\bf{w}}_{lk}|^2}\over{{\sum\nolimits_{i\ne
k}|\sum_{l\in\mathcal{A}}{\bf{h}}_{kl}^{\sf{H}}{\bf{w}}_{li}|^2}}+\sigma_{k}^2},
\forall k\in\mathcal{S}.
\end{eqnarray}
Each RRH has its own transmit power constraint
\begin{eqnarray}
\label{powerconstraint1}
\sum\limits_{k=1}^{K}\|{\bf{w}}_{lk}\|_{\ell_{2}}^2\le P_{l}, \forall l\in\mathcal{A}.
\end{eqnarray}

\subsection{Power Model}
The network power model is critical for the investigation of
 the energy efficiency of Cloud-RAN, which is described as follows. 

\subsubsection{RRH Power Consumption Model}
We will adopt the  following empirical linear model \cite{Auer_WC2011} for the power consumption of an RRH:
\begin{eqnarray}
P_{l}^{\textrm{rrh}} = \left\{ \begin{array}{ll}
 P_{a,l}^{\textrm{rrh}}+{1\over{\eta_{l}}} P_{l}^{\textrm{out}}, & \textrm{if $P_{l}^{\textrm{out}}>0$},\\
 P_{s,l}^{\textrm{rrh}}, & \textrm{if $P_{l}^{\textrm{out}}=0$}.
  \end{array} \right.
\end{eqnarray}
where $P_{a,l}^{\rm{rrh}}$ is the active power consumption, which depends on the number of antennas $N_{l}$, $P_{s,l}^{\rm{rrh}}$ is  the power consumption in the sleep mode, $P^{\textrm{out}}$ is the transmit power,
and $\eta_{l}$ is the drain efficiency of the radio frequency (RF) power amplifier. For the Pico-BS, the typical values are $P_{a,l}^{\textrm{rrh}}=6.8 W$, $P_{s,l}^{\textrm{rrh}}=4.3W$, and $\eta_{l}=4$ \cite{Auer_WC2011}. Based on this power consumption model, we conclude that it is essential to put the RRHs into sleep if possible.

\subsubsection{Transport  Network Power Consumption Model}
Although there is no superior solution to meet the low-cost, high-bandwidth, low-latency requirement of transport networks for the Cloud-RAN, the future passive optical network (PON)  can provide cost-effective connections between the RRHs and the BBU pool \cite{Kani_ICP2012}. PON comprises an optical line terminal (OLT) that connects a set of associated optical network units (ONUs) through a single fiber. Implementing a sleep mode in the optical network unit
(ONU)
has been considered as the most cost-effective and promising power-saving method  \cite{Dhaini_ITN2013} for the PON, but the OLT cannot go into the sleep mode and its power consumption is fixed\cite{Dhaini_ITN2013}. Hence, the total power consumption of the transport network is given by \cite{Dhaini_ITN2013}
\begin{eqnarray}
P^{\textrm{tn}}=P_{\textrm{olt}}+\sum_{l=1}^{L}P_{l}^{\textrm{tl}},
\end{eqnarray}
where $P_{\textrm{olt}}$ is the OLT power consumption, $P_{l}^{\textrm{tl}}=P_{a,l}^{\textrm{tl}}$ and $P_{l}^{\textrm{tl}}=P_{s,l}^{\textrm{tl}}$ denote the power consumed by the ONU $l$ (or the transport link $l$) in the active mode and sleep mode, respectively. The typical values are $P_{\textrm{olt}}=20 W$, $P_{a,l}^{\textrm{tl}}=3.85W$ and $P_{s,l}^{\textrm{tl}}=0.75W$ \cite{Dhaini_ITN2013}. Thus, we conclude that putting some transport links into the sleep   mode is a promising way to reduce the power consumption of Cloud-RAN. 
\subsubsection{Network Power Consumption}
Based on the above discussion, we can define a network power consumption model for the Cloud-RAN. Define $P_{l}^{a}=P_{a,l}^{\textrm{tl}}+P_{a,l}^{\textrm{tl}}$ $(P_{l}^{s}=P_{s,l}^{\textrm{rrh}}+P_{s,l}^{\textrm{tl}})$ as the active (sleep) power consumption when both the RRH and the corresponding transport link are switched on (off). For convenience, denote $P_{l}^c=P_{l}^a-P_l^s$. In the following, we will omit the constants  $P_{\textrm{olt}}$ and $\sum_{l=1}^{L}P_{l}^s$  , which will not affect the system design. Thus, the network power consumption is given by\begin{eqnarray}
\label{power_con1}
p(\mathcal{A},{\bf{w}})=\sum\limits_{l\in\mathcal{A}}\sum\limits_{k=1}^{K}{1\over{\eta_{l}}}\|{\bf{w}}_{lk}\|_{\ell_2}^2+\sum\limits_{l\in\mathcal{A}}P_{l}^{c},
\end{eqnarray}
where ${\bf{w}}=[{\bf{w}}_{11}^{T},\dots,{\bf{w}}_{1K}^{T},\dots,{\bf{w}}_{L1}^{T},\dots,{\bf{w}}_{LK}^{T}]^{T}$. In the following discussion, we shall refer to $P_{l}^c$ as the \emph{transport link power consumption} for simplification. Therefore, the first part of (\ref{power_con1}) is the total transmit power consumption and the second part is the total transport network power consumption. 

\section{Problem Formulation and Analysis}
Based on the power model, we will formulate the network power consumption minimization problem in this section.
\subsection{Power Saving Strategies and Problem Formulation}
The network power consumption model $(\ref{power_con1})$ indicates the following two strategies to reduce the network power consumption: 
\begin{itemize}
\item  Reduce the transmission power consumption;
\item Reduce the number of active RRHs and the corresponding transport links.
\end{itemize}
However, the two strategies conflict with each other. Specifically, in order to reduce the transmission power consumption, more RRHs are required to be active to exploit a higher beamforming gain. On the other hand, allowing more RRHs to be active will increase the power consumption of transport links. As a result, the network power consumption minimization problem requires a joint design of RRH (and the corresponding transport link) selection and coordinated transmit beamforming.

In this work, we assume perfect channel state information (CSI) available at the BBU pool.  With target SINRs  ${\boldsymbol{\gamma}}=(\gamma_{1},\dots,\gamma_{K})$, the network power consumption minimization problem can be formulated as
\begin{eqnarray}
\label{global1}
\mathscr{P}:
\mathop {\rm{minimize}}_{\{{\bf{w}}_{lk}\}, \mathcal{A}}&& p(\mathcal{A}, {\bf{w}})\nonumber\\
\rm{subject~to}&& {{|\sum_{l\in\mathcal{A}}{\bf{h}}_{kl}^{\sf{H}}{\bf{w}}_{lk}|^2}\over{{\sum\nolimits_{i\ne
k}|\sum_{l\in\mathcal{A}}{\bf{h}}_{kl}^{\sf{H}}{\bf{w}}_{li}|^2}}+\sigma_{k}^2}\ge\gamma_{k},
\nonumber\\
&&\sum\nolimits_{k=1}^{K}\|{\bf{w}}_{lk}\|_{\ell_2}^2\le P_{l}, l\in\mathcal{A}.
\end{eqnarray} 
Problem $\mathscr{P}$ is a joint RRH set selection and transmit beamforming problem, which is difficult to solve in general. In the following, we will analyze and reformulate it.

\subsection{Problem Analysis}
We first consider the case with a given active RRH set $\mathcal{A}$ for problem $\mathscr{P}$, resulting a network power minimization problem $\mathscr{P}(\mathcal{A})$. Let ${\bf{w}}_{k}=[{\bf{w}}_{lk}^{T}]^{T}\in\mathbb{C}^{\sum_{l\in\mathcal{A}}N_{l}}$ indexed by ${l\in\mathcal{A}}$, and ${\bf{h}}_{k}=[{\bf{h}}_{lk}^{T}]^{T}\in\mathbb{C}^{\sum_{l\in\mathcal{A}}N_{l}}$ indexed by ${l\in\mathcal{A}}$, such that ${\bf{h}}_{k}^{\sf{H}}{\bf{w}}_{k}=\sum_{l\in\mathcal{A}}{\bf{h}}_{kl}^{\sf{H}}{\bf{w}}_{lk}$. Since the phases of ${\bf{w}}_{k}$ will not change the objective function and constraints of $\mathscr{P}(\mathcal{A})$ \cite{Shamai_SP2006}, the SINR constraints are equivalent to the following second order cone (SOC) constraints:
\begin{eqnarray}
\label{SOC1}
\mathcal{C}_{1}(\mathcal{A}):
\sqrt{\sum\nolimits_{i\ne k}|{\bf{h}}_{k}^{\sf{H}}{\bf{w}}_{i}|^2+\sigma_{k}^2}\le {1\over{\sqrt{\gamma_{k}}}}\mathfrak{R}({\bf{h}}_{k}^{\sf{H}}{\bf{w}}_{k}), k\in\mathcal{S}. 
\end{eqnarray}
The per-RRH power constraints (\ref{powerconstraint1}) can be rewritten as
\begin{eqnarray}
\label{SOC2}
\mathcal{C}_{2}(\mathcal{A}): 
\sqrt{\sum\nolimits_{k=1}^{K}\|{\bf{A}}_{lk}{\bf{w}}_{k}\|_{\ell_2}^2}\le \sqrt{P_{l}}, l\in\mathcal{A},
\end{eqnarray}
where ${\bf{A}}_{lk}\in\mathbb{C}^{\sum_{l\in\mathcal{A}}N_{l}\times \sum_{l\in\mathcal{A}}N_{l}}$ is a block diagonal matrix with the identity matrix ${\bf{I}}_{N_{l}}$ as the $l$-th main diagonal block square matrix and zeros elsewhere.
Therefore, given the active RRH set $\mathcal{A}$, the network power minimization problem is given by
\begin{eqnarray}
\label{op1}
\mathscr{P}(\mathcal{A}): \mathop {\rm{minimize}}\limits_{{\bf{w}}_{1},\dots, {\bf{w}}_{K}} &&\sum_{l\in\mathcal{A}}\left(\sum_{k=1}^{K}{1\over{\eta_{l}}}\|{\bf{A}}_{lk}{\bf{w}}_{k}\|_{\ell_2}^2+P_{l}^c\right)\nonumber\\
{\rm{subject~to}} && \mathcal{C}_{1}(\mathcal{A}), \mathcal{C}_{2}(\mathcal{A}), 
\end{eqnarray}
with the optimal value denoted as $p^{\star}(\mathcal{A})$. This is a second-order cone programming (SOCP) problem, and can be solved efficiently, e.g., via interior point methods \cite{boyd2004convex}. 

Based on the solution of $\mathscr{P}(\mathcal{A})$, the network power minimization problem $\mathscr{P}$ can
be solved by searching over all the possible RRH sets,
i.e., \begin{eqnarray}
\label{exhaustive1}
p^{\star}=\mathop{\rm{minimize}}_{Q\in\{J, \dots, L\}}~p^{\star}(Q),
\end{eqnarray}
where $J\ge1$ is the minimum number of RRHs that makes the network support the QoS requirements, and $p^{\star}(Q)$ is determined by
\begin{eqnarray}
\label{greedy_1}
p^{\star}(Q)=\mathop {\rm{minimize}}_{\mathcal{A}\subseteq\mathcal{L}, |\mathcal{A}|=Q}~p^{\star}({\mathcal{A}}),
\end{eqnarray}
where $p^{\star}(\mathcal{A})$ is the optimal value of the problem $\mathscr{P}(\mathcal{A})$
(\ref{op1}) and $|\mathcal{A}|$ is the cardinality of set $\mathcal{A}$.
The number of subsets $\mathcal{A}$ of size $m$ is $L \choose m$, which can
be very large. Thus, in general, the overall procedure will be exponential
in the number of RRHs $L$ and thus cannot be applied in practice. Therefore, we will reformulate this problem to develop more efficient algorithms to solve it.

\subsection{Group Sparse Beamforming Formulation}
One way to solve problem $\mathscr{P}$ is to reformulate it as a MINLP problem \cite{Yuanming_Globecom2013}, and the generic algorithms for solving MINLP can be applied. Unfortunately, due to the high complexity, such an approach can only provide a performance benchmark for a simple network setting. In the following, we will pursue a different approach, and try to exploit the problem structure.

We will exploit the group sparsity of the optimal aggregative beamforming vector ${\bf{w}}$, which can be written as a partition:
\begin{eqnarray}
\label{vector1}
{\bf{w}}=[\underbrace{{\bf{w}}_{11}^{T},\dots,{\bf{w}}_{1K}^{T}}_{\tilde{\bf{w}}_{1}^{T}},\dots,\underbrace{{\bf{w}}_{L1}^{T},\dots,{\bf{w}}_{LK}^{T}}_{\tilde{\bf{w}}_{L}^{T}}]^{T},
\end{eqnarray} 
where all the coefficients in a given vector $\tilde{\bf{w}}_{l}=[{\bf{w}}_{l1}^{T},\dots,{\bf{w}}_{lK}^{T}]^{T}\in\mathbb{C}^{KN_{l}}$ form a group. When the RRH $l$ is switched off, the corresponding coefficients in the vector $\tilde{\bf{w}}_{l}$ will be set to zeros simultaneously. Overall there may be multiple RRHs switched off and the corresponding beamforming vectors will be set to zeros. That is, ${\bf{w}}$ has a group sparsity structure, with the priori knowledge that the blocks of variables in $\tilde{\bf{w}}_{l}$'s should be selected (the corresponding RRH will be switched on) or ignored (the corresponding RRH will be switched off)
simultaneously. 

Define $N=K\sum_{l=1}^{L}N_{l}$ and an index set $\mathcal{V}=\{1,2,\dots,N\}$ with its power-set as $2^{\mathcal{V}}=\{\mathcal{I}, \mathcal{I}\subseteq\mathcal{V}\}$. Furthermore, define the sets $\mathcal{G}_{l}=\{K\sum_{i=1}^{l-1}N_{i}+1,\dots,K\sum_{i=1}^{l}N_{i}\}, l=1,\dots, L$, as a partition of $\mathcal{V}$, such that $\tilde{\bf{w}}_{l}=[w_{i}]$ is indexed by $i\in\mathcal{G}_{l}$. Define the support of beamformer $\bf{w}$ as
\begin{eqnarray}
\mathcal{T}({\bf{w}})=\{i, w_{i}\ne0\},
\end{eqnarray}
where ${\bf{w}}=[w_{i}]$ is indexed by $i\in\mathcal{V}$. Hence, the transport link power consumption can be written as
\begin{eqnarray}
\label{transportnetwork1}
F(\mathcal{T}({\bf{w}}))=\sum\limits_{l=1}^{L}P_{l}^c I(\mathcal{T}({\bf{w}})\cap\mathcal{G}_{l}\ne\emptyset),
\end{eqnarray}
where $I(\mathcal{T}\cap\mathcal{G}_{l}\ne\emptyset)$ is an indicator function that takes
value 1 if $\mathcal{T}\cap\mathcal{G}_{l}\ne\emptyset$ and 0 otherwise. Therefore, the network power minimization problem $\mathscr{P}$  is equivalent to the following  group sparse beamforming (GSBF) formulation \begin{eqnarray}
\label{networkpower1}
\mathscr{P}_{\textrm{sparse}}: 
\mathop {\rm{minimize}}_{\bf{w}}\!\!\!\!\!\!\!&&T({\bf{w}})+F(\mathcal{T}(
{\bf{w}}))\nonumber\\
{\rm{subject~ to}}\!\!\!\!\!\!\! && \mathcal{C}_{1}(\mathcal{L}), \mathcal{C}_{2}(\mathcal{L}),
\end{eqnarray}
where $T({\bf{w}})=\sum_{l=1}^{L}\sum_{k=1}^{K}{1\over{\eta_{l}}}\|{\bf{w}}_{lk}\|_{\ell_2}^2$
represents the total transmit power consumption. The equivalence means that if ${\bf{w}}^{\star}$ is a solution to $\mathscr{P}_{\textrm{sparse}}$,
then $(\{{\bf{w}}_{lk}^{\star}\}, \mathcal{A}^{\star})$ with $\mathcal{A}^{\star}=\{l:\mathcal{T}({\bf{w}}^{\star})\cap\mathcal{G}_{l}\ne\emptyset\}$ is a solution to $\mathscr{P}$,
and vice versa. 

Note that the group sparsity of ${\bf{w}}$ is fundamentally different from the conventional sparsity measured by the $\ell_{0}$-norm of ${\bf{w}}$, which is often used in compressed sensing \cite{Tao_IT06,Donoho_TIT2006}. The reason is that although the $\ell_{0}$-norm of ${\bf{w}}$ will result in a sparse solution for ${\bf{w}}$, the zero entries of ${\bf{w}}$ will not necessarily align to a same group $\tilde{\bf{w}}_{l}$ to lead to switch off one RRH. As a result, the conventional $\ell_{1}$-norm relaxation \cite{Tao_IT06,Donoho_TIT2006} to the $\ell_{0}$-norm will not work for our problem due to the group sparsity of ${\bf{w}}$. Therefore, we will adopt the mixed $\ell_{1}/\ell_{p}$-norm \cite{Bach_ML2011} to induce group sparsity for ${\bf{w}}$. The details will be presented in Section V. 
 
Since obtaining the global optimization solutions to problem $\mathscr{P}$ is computationally difficult, in the following sections, we will propose two low-complexity algorithms to solve it. We will first propose a greedy algorithm in Section IV, which can be viewed as an approximation to the iteration procedure of (\ref{exhaustive1}). In order to further reduce the complexity, based on the GSBF formulation $\mathscr{P}_{\textrm{sparse}}$, a three-stage GSBF framework will then be developed based on the group-sparsity inducing norm minimization in Section V. 
\section{Greedy Selection Algorithm}
In this section, we develop a heuristic algorithm to solve $\mathscr{P}$ based on the backward greedy selection, which was successfully applied in spare filter design \cite{Oppenheim_SP2010} and has been shown to often yield optimal or near-optimal solutions. The backward greedy selection algorithm iteratively selects one RRH  to switch off at each step, while re-optimizing the coordinated transmit beamforming for the remaining active RRH set. The key design element for this algorithm is the selection rule of the RRHs to determine which one should be switched off at each step.

\subsection {Greedy Selection Procedure}
Denote the iteration number as $i=0,1,2,\dots$. At the $i$th iteration, $\mathcal{A}^{[i]}\subseteq\mathcal{L}$
shall denote the set of active RRHs, and $\mathcal{Z}^{[i]}$ denotes the inactive RRH set with $\mathcal{Z}^{[i]}\cup\mathcal{A}^{[i]}=\mathcal{L}$. In iteration $i$, an additional RRH $r^{[i]}\in\mathcal{A}^{[i]}$
will be added to $\mathcal{Z}^{[i]}$, resulting in a new set $\mathcal{Z}^{[i+1]}=\mathcal{Z}^{[i]}\cup\{r^{[i]}\}$
after this iteration. We initialize by setting $\mathcal{Z}^{[0]}=\emptyset$. In our algorithm, once an RRH is added to the set $\mathcal{Z}$, it cannot
be removed. This procedure is a simplification of the exact search method described in Section III-B. At iteration $i$, we need to solve the network power minimization
problem $\mathscr{P}(\mathcal{A}^{[i]})$ (\ref{op1}) with the given active RRH set $\mathcal{A}^{[i]}$.

\subsubsection{RRH Selection Rule}
How to select $r^{[i]}$ at the $i$th iteration is critical for the performance of the greedy selection algorithm. Based on our objective, we propose to select $r^{[i]}$ to 
maximize the decrease in the network power consumption. Specifically, at iteration $i$, we obtain the network power consumption $p^{\star}(\mathcal{A}_{m}^{[i]})$ with $\mathcal{A}_{m}^{[i]}\cup\{m\}=\mathcal{A}^{[i]}$ by removing any $m\in\mathcal{A}^{[i]}$ from the active RRH set $\mathcal{A}^{[i]}$. Thereafter, $r^{[i]}$ is chosen to yield the smallest network power consumption after switching off the corresponding RRH, i.e.,   
\begin{eqnarray}
\label{selection1}
r^{[i]}=\arg\min_{m\in\mathcal{A}^{[i]}} p^{\star}(\mathcal{A}_{m}^{[i]}).
\end{eqnarray}
We assume that $p^{\star}(\mathcal{A}_{m}^{[i]})=+\infty$ if problem $\mathscr{P}(\mathcal{A}_{m}^{[i]})$ is infeasible. The impact
of switching off one RRH is reducing the transport
network power consumption while increasing the total transmit power consumption. Thus, the proposed selection rule actually aims at minimizing the impact of turning off one RRH at each iteration.

Denote $\mathcal{J}$ as the set of candidate RRHs that can be turned off, the greedy selection algorithm is described as follows:

\begin{algorithm}
\caption{Greedy Selection Algorithm }
\textbf{Step 0:} Initialize $\mathcal{Z}^{[0]}=\emptyset$, $\mathcal{A}^{[0]}=\{1,\dots,L\}$
and $i=0$;\\
\textbf{Step 1:} Solve the optimization problem $\mathscr{P}(\mathcal{A}^{[i]})$
(\ref{op1});\\
\begin{enumerate}
\item {\bf{If}} (\ref{op1}) is feasible, obtain $p^{\star}(\mathcal{A}^{[i]})$;\\
\begin{itemize}
\item {\bf{If}} $\forall m\in\mathcal{A}^{[i]}$, problem $\mathscr{P}(\mathcal{A}_{m}^{[i]})$ is infeasible, obtain $\mathcal{J}=\{0,\dots, i\}$, {\textbf{go to Step 2}};\\
\item {\bf{If}} $\exists m\in\mathcal{A}^{[i]}$, make problem $\mathscr{P}(\mathcal{A}_{m}^{[i]})$
feasible, find the $r^{[i]}$ according to (\ref{selection1}) and update the set $\mathcal{Z}^{[i+1]}=\mathcal{Z}^{[i]}\cup\{r^{[i]}\}$ and the iteration number $i\leftarrow i+1$, \textbf{go to Step 1};
\end{itemize}
\item \textbf{If} (\ref{op1}) is infeasible, when $i=0$, $p^{\star}=\infty$, \textbf{go
to End}; when $i>0$, obtain $\mathcal{J}=\{0,1, \dots, i-1\}$,\\ \textbf{go to Step 2};
\end{enumerate}
\textbf{Step 2:} Obtain the optimal active RRH set $\mathcal{A}^{[j^{\star}]}$
with $j^{\star}=\arg\min_{j\in\mathcal{J}} p^{\star}(\mathcal{A}^{[j]})$ and the transmit beamformers minimizing $\mathscr{P}(\mathcal{A}^{[j^{\star}]})$;\\
\textbf{End}
\end{algorithm}

\subsection{Complexity Analysis}
At the $i$-th iteration, we need to solve $|\mathcal{A}^{[i]}|$ SCOP problems $\mathscr{P}({\mathcal{A}}_{m}^{[i]})$ by removing the RRH $m$ from the set $\mathcal{A}^{[i]}$ to determine which RRH should be selected. For each of the SOCP problem $\mathscr{P}(\mathcal{A})$, using the interior-point method, the computational complexity is $\mathcal{O}((K\sum_{l\in\mathcal{A}}N_{l})^{3.5})$ \cite{boyd2004convex}. The total number of iterations is bounded by $L$. As a result, the total number of SOCP problems required to be solved grows \emph{quadratically} with $L$. Although this reduces the computational complexity significantly compared with the mixed-integer conic programming based algorithms in \cite{leyffer_2012mixed} and \cite{Cheng_SP2013},   the complexity is still prohibitive for large-scale networks. Therefore, in the next section we will propose  a group sparse beamforming framework to further reduce the complexity. 

\section{Group Sparse Beamforming Framework}
In this section, we will develop two low-complexity algorithms based on  the GSBF formulation $\mathscr{P}_{\textrm{sparse}}$, namely a bi-section GSBF algorithm and  an iterative GSBF algorithm, for which, the overall number of SOCP problems to solve grows \emph{logarithmically} and \emph{linearly} with $L$, respectively. The main motivation is to induce group sparsity in the beamformer, which corresponds to switching off RRHs. 

In the bi-section GSBF algorithm, we will minimize the {\emph{weighted}} mixed $\ell_{1}/\ell_{2}$-norm to induce
group-sparsity for the beamformer. By exploiting the additional prior information (i.e., power amplifier efficiency, transport link power consumption, channel power gain) that available in our setting, the proposed bi-section GSBF algorithm will be demonstrated through
rigorous analysis and simulations to outperform the conventional \emph{unweighted} mixed $\ell_{1}/\ell_{p}$-norm minimization
substantially\cite{Z.Q.Luo_JSAC2013,TonyQ.S._WC2013,Mehanna_SP2013}. By minimizing the \emph{re-weighted} mixed $\ell_{1}/\ell_{2}$-norm iteratively to enhance the group sparsity for the beamformer, the proposed iterative GSBF algorithm will further improve the performance. 

\begin{figure}[h]
\centering
\includegraphics[width=1\columnwidth]{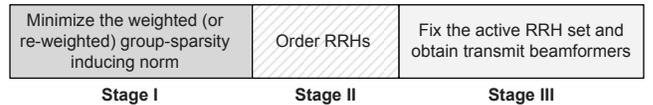}
\caption{A Three-Stage GSBF Framework.}
\label{GSBF}
\end{figure}

The proposed GSBF framework is a three-stage approach, as shown in Fig. {\ref{GSBF}}. Specifically, in the first stage, we minimize a weighted (or re-weighted) group-sparsity inducing norm to induce the group-sparsity in the beamformer. In the second stage, we propose an ordering rule to determine the priority for the RRHs that should be switched off, based on not only the (approximately)
sparse beamformer obtained in the first stage, but also some key system parameters. Following the ordering rule, a selection procedure is performed to determine the optimal active RRH set,  followed by the coordinated beamforming. The details will be presented in the following subsections.

\subsection{Preliminaries on Group-Sparsity Inducing Norms}   
The mixed $\ell_{1}/\ell_{p}$-norm has recently received lots of attention
and is shown to be effective to induce group sparsity \cite{Bach_ML2011}, which is defined as
follows:
\begin{definition} Consider the vector ${\bf{w}}=[{\bf{w}}_{lk}]$ indexed
by $l\in\mathcal{L}$ and $k\in\mathcal{S}$ as define in (\ref{vector1}).
Its mixed $\ell_{1}/\ell_{p}$-norm is defined as follows:
\begin{eqnarray}
\label{mixed1}
\mathcal{R}({\bf{w}})=\sum\limits_{l=1}^{L}\beta_{l}\|\tilde{\bf{w}}_{l}\|_{\ell_{p}},
~p>1,
\end{eqnarray}
\end{definition}
where $\beta_{1}, \beta_{2},\dots, \beta_{L}$ are positive weights.

Define the vector ${\bf{r}}= [\|\tilde{\bf{w}}_{1}\|_{\ell_p}, \dots, \|\tilde{\bf{w}}_{L}\|_{\ell_p}]^{T}$,
then the mixed $\ell_{1}/\ell_{p}$-norm behaves as the $\ell_{1}$-norm
on the vector ${\bf{r}}$, and therefore, inducing group sparsity (i.e., each vector
$\tilde{\bf{w}}_{l}$ is encouraged to be set to zero) for ${\bf{w}}$. Note
that, within the group $\tilde{\bf{w}}_{l}$, the $\ell_{p}$-norm does not
promote sparsity as $p>1$. By setting $p=1$, the mixed
$\ell_{1}/\ell_{p}$-norm becomes a weighted $\ell_{1}$-norm, which will not
promote group sparsity. The mixed $\ell_{1}/\ell_{2}$-norm and $\ell_{1}/\ell_{\infty}$-norm
are two commonly used norms for inducing group sparsity. For instance, the
mixed
$\ell_{1}/\ell_{2}$-norm is used with the name \emph{group least-absolute
selection} and \emph{shrinkage operator} (or \emph{Group-Lasso}) in machine
learning \cite{Ming_SM06}. In high dimensional statistics, the mixed
$\ell_{1}/\ell_{\infty}$-norm is adopted as a regularizer in the linear regression
problems with sparsity constraints for its computational convenience \cite{Wainwright_IT2011}. 

\subsection{Bi-Section GSBF Algorithm}
In this section, we propose a binary search based GSBF algorithm, in which, the overall number of SOCP problems required to be solved grows logarithmically with $L$, instead of quadratically for the GS algorithm.

\subsubsection{Group-Sparsity Inducing Norm Minimization}
With the combinatorial function $F(\cdot)$ in the\ objective function $p({\bf{w}})=T({\bf{w}})+F(\mathcal{T}(\bf{w}))$, the problem $\mathscr{P}_{\textrm{sparse}}$ becomes computationally intractable. Therefore, we first construct an appropriate convex relaxation for the objective function $p(\bf{w})$
as a surrogate objective function, resulting a weighted mixed $\ell_{1}/\ell_{2}$-norm
minimization problem to induce group sparsity for the beamformer. Specifically, we first derive its tightest positively homogeneous lower bound $p_{h}({\bf{w}})$, which has the property $p_{h}(\lambda {\bf{w}})=\lambda p_{h}({\bf{w}}), 0<\lambda< \infty$.  Since $p_{h}(\bf{w})$ is still not  convex, we further calculate its Fenchel-Legendre biconjugate $p_{h}^{**}({\bf{w}})$ to provide a tightest convex lower bound for $p_{h}(\bf{w})$. We call $p_{h}^{**}({\bf{w}})$ as the \emph{convex positively homogeneous lower bound} (the details can be found in \cite{Obozinski_arXiv2012}) of function $p(\bf{w})$, which is provided in the following theorem:

\begin{proposition}
The tightest convex positively homogeneous lower bound of  the objective
function in $\mathscr{P}_{\textrm{sparse}}$, denoted as $p({\bf{w}})$, is given by
\begin{eqnarray}
\label{plower}
\Omega({\bf{w}})=2\sum_{l=1}^{L}\sqrt{P_{l}^c\over{\eta_{l}}}\|\tilde{\bf{w}}_{l}\|_{\ell_2}.
\end{eqnarray}
\end{proposition}
\begin{IEEEproof}
Please refer to Appendix A. 
\end{IEEEproof}

This theorem indicates that the group-sparsity inducing norm (i.e., the weighted mixed
$\ell_{1}/\ell_{2}$-norm) can provide a convex relaxation for the objective
function $p(\bf{w})$. Furthermore, it encapsulates the additionally prior information in terms of system parameters into the weights for the groups. Intuitively, the weights indicate that the RRHs with a higher transport link power consumption and lower power amplifier efficiency will have a higher chance being forced to be switched off. Using the weighted mixed
$\ell_{1}/\ell_{2}$-norm as a surrogate for the objective function,
we minimize the weighted mixed $\ell_{1}/\ell_{2}$-norm $\Omega({\bf{w}})$ to induce the group-sparsity for the beamformer
${\bf{w}}$:
\begin{eqnarray}
\label{sp1}
\mathscr{P}_{\textrm{GSBF}}: \mathop {\rm{minimize}}_{\bf{w}} \!\!\!\!\!\!\!&&\Omega({\bf{w}})\nonumber\\
{\rm{subject~ to}}\!\!\!\!\!\!\! && \mathcal{C}_{1}(\mathcal{L}), \mathcal{C}_{2}(\mathcal{L}),
\end{eqnarray}
which is an SOCP problem and can be solved efficiently.

\subsubsection{RRH Ordering}
After obtaining the (approximately) sparse  beamformer $\hat{{\bf{w}}}$ via solving the weighted group-sparsity inducing norm minimization problem $\mathscr{P}_{\textrm{GSBF}}$, the next question is how to determine the active RRH set. We will first give priorities to different RRHs, so that an RRH with a higher priority should be switched off before the one with a lower priority. Most previous works \cite{Z.Q.Luo_JSAC2013,TonyQ.S._WC2013,Mehanna_SP2013} applying the idea of group-sparsity inducing norm minimization directly map the sparsity to their application, e.g., in \cite{Mehanna_SP2013}, the transmit antennas corresponding to the smaller coefficients in the group (measured by the $\ell_{\infty}$-norm) will have a higher priority to be switched off. In our setting, one might be tempted to give a higher priority for an RRH $l$ with a smaller coefficient $r_{l}=(\sum_{k=1}^{K}\|\hat{\bf{w}}_{lk}\|_{\ell_{2}}^2)^{1/2}$, as the RRH $l$ with a smaller coefficient $r_{l}$ may provide a lower beamforming gain and should be encouraged to be turned off. It turns out that such ordering rule is not a good option and will bring performance degradation.

To get a better performance, the priority of the RRHs should be determined
by not only the beamforming gain but also other key system parameters that indicate the impact of the RRHs on the network performance.
In particular, the channel power gain $\kappa_{l}=\sum\nolimits_{k=1}^{K}\|{\bf{h}}_{kl}\|_{\ell_2}^2$
should be taken into consideration.  Specifically, by the broadcast channel (BC)-multiple-access channel (MAC) duality \cite{Goldsmith_TIT2003}, we have the sum capacity of the Cloud-RAN as:
\begin{eqnarray}
C_{\textrm{sum}}=\log\det({\bf{I}}_{N}+{\sf{snr}}\sum_{k=1}^{K}{\bf{h}}_{k}{\bf{h}}_{k}^{\sf{H}}),
\end{eqnarray}
where we assume equal power allocation to simplify the analysis, i.e., ${\sf{snr}}=P/\sigma^2, \forall k=1,\dots,K$. One way to upper-bound $C_{\textrm{sum}}$ is through upper-bounding the capacity by the total receive SNR, i.e., using the following relation
\begin{eqnarray}
\log\det({\bf{I}}_{N}+{\sf{snr}}\sum_{k=1}^{K}{\bf{h}}_{k}{\bf{h}}_{k}^{\sf{H}})&\le&{\rm{Tr}}({\sf{snr}}\sum_{k=1}^{K}{\bf{h}}_{k}{\bf{h}}_{k}^{\sf{H}})\nonumber\\
&=&{\sf{snr}}\sum_{l=1}^{L}\kappa_{l},
\end{eqnarray}
which relies on the inequality $\log(1+x)\le x$.
Therefore, from the capacity perspective, the RRH with a higher channel power gain $\kappa_{l}$ contributes more to the sum capacity, i.e., it provides a higher power gain and should not be encouraged to be switched off.  

Therefore, different from the previous democratic
assumptions (e.g.,  \cite{Z.Q.Luo_JSAC2013,TonyQ.S._WC2013,Mehanna_SP2013}) on the mapping between the sparsity and their applications directly, we exploit the prior information in terms of system parameters to refine the mapping on the group-sparsity. Specifically, considering the key system parameters, we propose the following ordering criterion to determine which RRHs should be switched off, i.e.,
\begin{eqnarray}
\label{sparse_selection1}
\theta_{l}:=\sqrt{{\kappa_{l}\eta_{l}}\over{P_{l}^c}}\left(\sum_{k=1}^{K}\|\hat{\bf{w}}_{lk}\|_{\ell_{2}}\right)^{1/2}, \forall l=1,\dots,L,
\end{eqnarray} 
where the RRH with a smaller $\theta_{l}$ will have a higher priority to be switched off.
This ordering rule indicates that the RRH with a lower beamforming gain, lower channel
power gain, lower power amplifier efficiency, and higher transport link power consumption should have a higher priority to be switched off. The proposed ordering rule will be demonstrated to improve the performance of the GSBF algorithm significantly through simulations.

\subsubsection{Binary Search Procedure}
Based on the ordering rule (\ref{sparse_selection1}), we sort the coefficients  in the ascending order: $\theta_{\pi_{1}}\le\theta_{\pi_{2}}\le\cdots\le\theta_{\pi_{L}}$ to fix the final active RRH set. We set the first $J$ smallest coefficients to zero, as a result, the corresponding RRHs will be turned. Denote $J_{0}$ as the maximum number of RRHs that can be turned off, i.e., the problem $\mathscr{P}(\mathcal{A}^{[i]})$ is infeasible if $i> J_{0}$, where $\mathcal{A}^{[i]}\cup\mathcal{Z}^{[i]}=\mathcal{L}$ with $\mathcal{Z}^{[i]}=\{{\pi_{0}},{\pi_{1}},\dots,{\pi_{i}}\}$ and $\pi_{0}=\emptyset$. A binary search procedure can be adopted to determine $J_{0}$, which only needs to solve no more than $1+\lceil\log(1+L)\rceil$ SOCP problems. In this algorithm, we regard $\mathcal{A}^{[J_{0}]}$ as the final active RRH set and the solution of $\mathscr{P}(\mathcal{A}^{[J_{0}]})$ is the final transmit beamformer.  

Therefore,
the bi-section GSBF algorithm is presented as follows: 
\begin{algorithm}
\caption{Bi-Section GSBF Algorithm }
\textbf{Step 0:} Solve the weighted group-sparsity inducing norm minimization problem $\mathscr{P}_{\textrm{GSBF}}$;\\
\begin{enumerate}
\item {\textbf{If}} it is infeasible, set $p^{\star}=\infty$, {\textbf{go to End}};
\item  {\textbf{If}} it is feasible, obtain the solution $\hat{\bf{w}}$,
calculate ordering criterion (\ref{sparse_selection1}), and sort them in the \\
ascending order: $\theta_{\pi_{1}}\le\dots\le\theta_{\pi_{L}}$, {\textbf{go to Step 1}};
\end{enumerate}
\textbf{Step 1:} Initialize $J_{\textrm{low}}=0$, $J_{\textrm{up}}=L$, $i=0$;\\
\textbf{Step 2:} Repeat
\begin{enumerate}
\item Set $i\leftarrow\lfloor{{J_{\textrm{low}}+J_{\textrm{up}}}\over{2}}\rfloor$;\\
\item Solve the optimization problem $\mathscr{P}(\mathcal{A}^{[i]})$ (\ref{op1}): if it is infeasible, set $ J_{\textrm{low}}=i$; otherwise, set $J_{\textrm{up}}=i$;
\end{enumerate}
\textbf{Step 3:} Until $J_{\textrm{up}}-J_{\textrm{low}}=1$, obtain $J_{0}=J_{\textrm{low}}$ and obtain the optimal active RRH set $\mathcal{A}^{\star}$ with $\mathcal{A}^{\star}\cup\mathcal{J}=\mathcal{L}$ and $\mathcal{J}=\{{\pi_{1}},\dots, {\pi_{J_{0}}}\}$;\\
\textbf{Step 4:} Solve the problem $\mathscr{P}(\mathcal{A}^{\star})$, obtain the minimum network power consumption and the corresponding transmit beamformers;\\
\textbf{End}
\end{algorithm}

\subsection{Iterative GSBF Algorithm}
Under the GSBF framework, the main task of the first two stages is to order the RRHs according to the criterion (\ref{sparse_selection1}), which depends on the sparse solution to $\mathscr{P}_{\textrm{GSBF}}$, i.e., $\{\hat{\bf{w}}_{lk}\}$. However, when the minimum of $r_{l}=(\sum_{k=1}^{K}\|\hat{\bf{w}}_{lk}\|_{\ell_{2}}^2)^{1/2}>0$ is not close to zero, it will introduce large bias in estimating which RRHs can be switched off. To resolve this issue, we will apply the idea from the  majorization-minimization (MM) algorithm \cite{hunter2004tutorial} (please refer to appendix B for details on this algorithm), to enhance group-sparsity for the beamformer to better estimate which RRHs can be switched off. 

The MM algorithms have been successfully applied in the re-weighted $\ell_{1}$-norm (or mixed $\ell_{1}/\ell_{2}$-norm) minimization problem to enhance sparsity \cite{Boyd_2008enhancing, TonyQ.S._WC2013,Mehanna_SP2013}. However, these algorithms failed to exploit the additional system prior information to improve the performance. Specifically, they used the un-weighted $\ell_{1}$-norm (or mixed $\ell_{1}/\ell_{p}$-norm) minimization as the start point of the iterative algorithms and re-weighted the $\ell_{1}$-norm (or mixed $\ell_{1}/\ell_{p}$-norm) only using the estimate of the coefficients obtained in the last minimization step.  
Different from the above conventional re-weighted algorithms, we exploit the additionally system prior information at each step (including the start step) to improve the estimation on the group-sparsity of the beamformer.

\subsubsection{Re-weighted Group-Sparsity Inducing Norm Minimization}
One way to enhance the group-sparsity compared with using the weighted mixed $\ell_{1}/\ell_{2}$ norm
$\Omega({\bf{w}})$ in (\ref{plower}) is to minimize the following combinatorial function directly:
\begin{eqnarray}
\label{enhancesparse}
\mathcal{R}({\bf{w}})=2\sum_{l=1}^{L}\sqrt{P_{l}^c\over{\eta_{l}}}I(\|\tilde{{\bf{w}}}_{l}\|_{\ell_2}>0),
\end{eqnarray}
for which the convex function $\Omega({\bf{w}})$ in (\ref{plower}) can be regarded as an $\ell_{1}$-norm relaxation. Unfortunately, minimizing $\mathcal{R}(\bf{w})$ will lead to a non-convex optimization problem. In this subsection, we will provide a sub-optimal algorithm to solve (25) by adopting the idea from the MM algorithm to enhance sparsity.

Based on the following fact in \cite{sriperumbudur2011majorization}
\begin{eqnarray}
\lim_{\epsilon\rightarrow0}{{{\rm{log}}(1+x\epsilon^{-1})}\over{\rm{log}}(1+\epsilon^{-1})}
= \left\{ \begin{array}{ll}
 0 & \textrm{if $x=0$},\\
 1 & \textrm{if $x>0$},
  \end{array} \right.
\end{eqnarray}
we rewrite the indicator function in (\ref{enhancesparse}) as
\begin{eqnarray}
\label{mm_power}
I(\|\tilde{{\bf{w}}_{l}}\|_{\ell_2}>0)=\lim_{\epsilon\rightarrow
0}{{{\rm{log}}(1+\|\tilde{\bf{w}}_{l}\|_{\ell_{2}}\epsilon^{-1})}\over{{\rm{log}}(1+\epsilon^{-1})}}, \forall l\in\mathcal{L}.
\end{eqnarray}
The surrogate objective function $\mathcal{R}({\bf{w}})$ can then be approximated
as 
\begin{eqnarray}
\label{mmo}
f({\bf{w}})=\lambda_{\epsilon}\sum_{l=1}^{L}\sqrt{P_{l}^{c}\over{\eta_{l}}}{\rm{log}}(1+\|\tilde{\bf{w}}_{l}\|_{\ell_{2}}\epsilon^{-1}),
\end{eqnarray}
by neglecting the limit in (\ref{mm_power}) and choosing an appropriate $\epsilon>0$, where $\lambda_{\epsilon}={2\over{\rm{log}}(1+\epsilon^{-1})}$.  Compared with $\Omega({\bf{w}})$ in (\ref{plower}), the log-sum penalty function $f({\bf{w}})$ has the potential to be much more sparsity-encouraging. The detail explanations can be found in \cite{Boyd_2008enhancing}.

Since ${\textrm{log}}(1+x),
x\ge0$, is a concave function, we can construct a majorization function for
$f$ by the first-order approximation of $\textrm{log}(1+\|\tilde{\bf{w}}_{l}\|_{\ell_{2}}\epsilon^{-1})$,
i.e., 
\begin{eqnarray}
f({\bf{w}})\le \lambda_{\epsilon} \sum_{l=1}^{L}\sqrt{P_{l}^{c}\over{\eta_{l}}}\left(\underbrace{{{\|\tilde{\bf{w}}_{l}\|_{\ell_{2}}}\over{\|\tilde{\bf{w}}_{l}^{[m]}\|_{\ell_{2}}+\epsilon}}+c({\bf{w}}^{[m]})}_{g({\bf{w}}|{\bf{w}}^{[m]})}\right),
\end{eqnarray} 
where ${\bf{w}}^{[m]}$ is the minimizer at the $(m-1)$-th iteration, and $c({\bf{w}}^{[m]})={\rm{log}}(1\!+\!\|\tilde{\bf{w}}_{l}^{[m]}\|_{\ell_{2}})\!-\!{{\|\tilde{\bf{w}}_{l}^{[m]}\|_{\ell_2}}/({\|\tilde{\bf{w}}_{l}^{[m]}\|_{\ell_{2}}\!+\!\epsilon}})$ is a constant provided that ${\bf{w}}^{[m]}$ is already known at the current $m$-th iteration.

By omitting the constant part of $g({\bf{w}}|{\bf{w}}^{[m]})$ at the $m$-th iteration, which will not affect the solution, we propose a re-weighted GSBF framework to enhance the group-sparsity:
\begin{eqnarray}
\label{wGSBF}
\!\!\!\!\!\!\!\!\!\! \mathscr{P}_{\textrm{iGSBF}}^{[m]}\!:\!\{\tilde{{\bf{w}}}_{l}^{[m+1]}\}_{l=1}^{L}\!=\!\arg\min\!\!\!\!\!\!\!\!\!\!\!&& \sum_{l=1}^{L}\beta_{l}^{[m]}\|\tilde{\bf{w}}_{l}\|_{\ell_{2}}\nonumber\\
{\rm{subject~to}}\!\!\!\!\!\!\!\!\!\! &&\mathcal{C}_{1}(\mathcal{L}), \mathcal{C}_{2}(\mathcal{L}),
\end{eqnarray}  
where 
\begin{eqnarray}
\label{upweights}
\beta_{l}^{[m]}=\sqrt{P_{l}^c\over{\eta_{l}}}{{1}\over{(\|\tilde{\bf{w}}_{l}^{[m]}\|_{\ell_{2}}+\epsilon)}},\forall l=1,\dots, L,
\end{eqnarray}
are the weights for the groups at the $m$-th iteration. At each step, the mixed $\ell_{1}/\ell_{2}$-norm optimization is re-weighted using the estimate of the beamformer obtained in the last minimization step.

As this iterative algorithm cannot guarantee the global minimum, it is important to choose a suitable starting point to obtain a good local optimum. As suggested in \cite{TonyQ.S._WC2013,Boyd_2008enhancing,Mehanna_SP2013},
this algorithm can be initiated with the solution of the unweighted $\ell_{1}$-norm
minimization, i.e., $\beta_{l}^{[0]}=1, \forall l=1,\dots,L$. In our
setting, however, the prior information on the system parameters can help us generate a high quality stating point for the iterative GSBF framework. Specifically, with the available channel state information, we choose the $\ell_{2}$-norm of the   initial beamformer at the $l$-th RRH $\|\tilde{{\bf{w}}}_{l}^{[0]}\|_{\ell_2}$ to be proportional to its corresponding channel power gain $\kappa_{l}$, arguing that the RRH with a low channel power gain should be encouraged to be switched off as justified in section V-B. Therefore, from (\ref{upweights}), we set the following weights as the initiation weights
for $\mathscr{P}_{\textrm{iGSBF}}^{[0]}$:
\begin{eqnarray}
\label{initiation1}
\beta_{l}^{[0]}=\sqrt{P_{l}^{c}\over{\eta_{l}\kappa_{l}}}, \forall l=1,\dots,L.
\end{eqnarray} 
The weights indicate that the RRHs with a higher transport
power consumption, lower power amplifier efficiency and lower channel power
gain should be penalized more heavily. 

As observed
in the simulations, this algorithm converges very fast (typically within
20 iterations). We set the maximum number of iteration as $m_{\textrm{max}}=L$ in our simulations.

\subsubsection{Iterative Search Procedure}
After obtaining the (approximately) sparse  beamformers using the above re-weighted GSBF framework, we still adopt the same ordering criterion (\ref{sparse_selection1})
to fix the final active RRH set. 

Different from the aggressive strategy in the bi-section GSBF algorithm, which assumes that the RRH should be switched off  as many as possible and thus results a minimum transport network power consumption, we adopt a conservative strategy to determine the final active RRH set by realizing that the minimum network power consumption may not be attained when the transport network power consumption is minimized. 

Specifically, denote $J_{0}$ as the maximum number of RRHs that can be turned off, the corresponding inactive RRH set is $\mathcal{J}=\{{\pi_{0}},{\pi_{1}},\dots,{\pi_{J_{0}}}\}$. The minimum network power consumption should be searched over all the values $\mathscr{P}^{*}(\mathcal{A}^{[i]})$, where $\mathcal{A}^{[i]}=\mathcal{L}\setminus\{\pi_{0},\pi_{1},\dots,\pi_{i}\}$ and $0\le i\le J_{0}$. This can be accomplished using an iterative search procedure that requires to solve no more than $L$ SOCP problems. 

Therefore, the overall iterative GSBF algorithm is presented as Algorithm 3:
\begin{algorithm}
\caption{Iterative GSBF Algorithm }
\textbf{Step 0:} Initialize the weights $\beta_{l}^{[0]}, l=1,\dots, L$ as in (\ref{initiation1}) and the iteration counter $m=0$;\\
\textbf{Step 1:} Solve the weighted GSBF problem $\mathscr{P}_{\textrm{iGSBF}}^{[m]}$ (\ref{wGSBF}): {\textbf{if}} it is infeasible, set $p^{\star}=\infty$ and {\textbf{go to End}}; otherwise, set $m=m+1$, {\textbf{go to Step 2}};\\
\textbf{Step 2:} Update the weights using (\ref{upweights}); \\
\textbf{Step 3:} {\textbf{If}} converge or $m=m_{\textrm{max}}$, obtain the solution $\hat{\bf{w}}$ and calculate the selection criterion (\ref{sparse_selection1}), and sort them \mbox{in the ascending order: $\theta_{\pi_{1}}\le\dots\le\theta_{\pi_{L}}$,  {\textbf{go to Step 4}}}; otherwise, \textbf{go to Step 1}; \\
\textbf{Step 4:} Initialize $\mathcal{Z}^{[0]}=\emptyset$, $\mathcal{A}^{[0]}=\{1,\dots,L\}$, and $i=0$;\\
\textbf{Step 5:} Solve the optimization problem $\mathscr{P}(\mathcal{A}^{[i]})$ (\ref{op1});
\begin{enumerate}
\item {\textbf{If}} (\ref{op1}) is feasible, obtain $p^{*}(\mathcal{A}^{[i]})$, update the set $\mathcal{Z}^{[i+1]}=\mathcal{Z}^{[i]}\cup\{\pi_{i+1}\}$ and $i=i+1$, {\textbf{go to Step 5}};
\item {\textbf{If}} (\ref{op1}) is infeasible, obtain $\mathcal{J}=\{0,1,\dots, i-1\}$, {\textbf{go to Step 6}};\\
\end{enumerate} 
{\textbf{Step 6:}} Obtain optimal RRH set $\mathcal{A}^{[j^{\star}]}$ and beamformers minimizing $\mathscr{P}(\mathcal{A}^{[j^{\star}]})$ with $j^{\star}=\arg\min_{j\in\mathcal{J}}p^{*}(\mathcal{A}^{[j]})$;\\
\textbf{End}
\end{algorithm}

\section{Simulation Results }

In this section, we simulate the performance of the proposed algorithms. We consider the following channel model
\begin{eqnarray}
{\bf{h}}_{kl}=10^{-L(d_{kl})/20}\sqrt{\varphi_{kl}s_{kl}}{\bf{g}}_{kl},
\end{eqnarray}
where $L(d_{kl})$ is the path-loss at distance $d_{kl}$, , as given in Table \ref{parameter1}, $s_{kl}$ is the shadowing coefficient, $\varphi_{kl}$ is the antenna gain and ${\bf{g}}_{kl}$ is the small scale fading coefficient. 
We use the standard cellular network parameters as showed in Table \ref{parameter1}. 
\begin{table}[t]
\renewcommand{\arraystretch}{1.3}
\caption{Simulation Parameters}
\label{table_example}
\centering
\begin{tabular}{l|c}
Parameter & Value\\
\hline
Path-loss at distance $d_{kl}$ (km) & 148.1+37.6 ${\textrm{log}}_{2}({d}_{kl})$
\\
Standard deviation of log-norm shadowing $\sigma_{s}$ & 8 dB\\
Small-scale fading distribution ${\bf{g}}_{kl}$ & $\mathcal{CN}({\bf{0}},
{\bf{I}})$\\
Noise power $\sigma_{k}^{2} $ \cite{Soliman_CM13} (10 MHz bandwidth)& -102
dBm\\
Maximum transmit power of RRH $P_{l}$ \cite{Soliman_CM13} & 1 W\\
Power amplifier efficiency $\alpha_{l}$ \cite{Auer_WC2011}& 4\\
Transmit antenna power gain & 9 dBi
\end{tabular}
\label{parameter1}
\end{table}
Each point of the simulation results
is averaged over 50 randomly generated network realizations. 

The proposed algorithms are compared to the following algorithms: 
\begin{itemize}
\item \textbf{Coordinated beamforming (CB) algorithm}: In this algorithm, all
the RRHs are active and only the total transmit power consumption is minimized \cite{WeiYu_WC10}.
\item \textbf{Mixed-integer nonlinear programming (MINLP) algorithm}: This algorithm \cite{leyffer_2012mixed,Cheng_SP2013} can obtain the global optimum. Since the complexity of the algorithm grows exponentially with the number of RRHs $L$, we only run it in a small-size  network.  
\item \textbf{Conventional Sparsity pattern (SP) based algorithm}: In this algorithm, the unweighted mixed $\ell_{1}/\ell_{p}$-norm is adopted to induce group sparsity as in \cite{Z.Q.Luo_JSAC2013} and \cite{Mehanna_SP2013}. The ordering of RRHs is determined only by the group-sparsity of the beamformer, i.e., $\theta_{l}:=(\sum_{k=1}^{K}\|\hat{\bf{w}}_{lk}\|_{\ell_{2}})^{1/2},
\forall l=1,\dots,L$, instead of (\ref{sparse_selection1}). The complexity of the algorithm grows logarithmically with $L$. 
 
\item \textbf{Relaxed mixed-integer nonlinear programming (RMINLP) based algorithm}: In this algorithm, a deflation procedure is performed to switch off RRHs one-by-one based on the solutions  obtained via solving the relaxed MINLP by relaxing the integers to the unit intervals \cite{Cheng_SP2013}. The complexity of the algorithm grows linearly with $L$.   
\end{itemize}
   
\subsection{Network Power Consumption versus Target SINR }
\begin{figure}[!t]
\centering
\includegraphics[width=1\columnwidth]{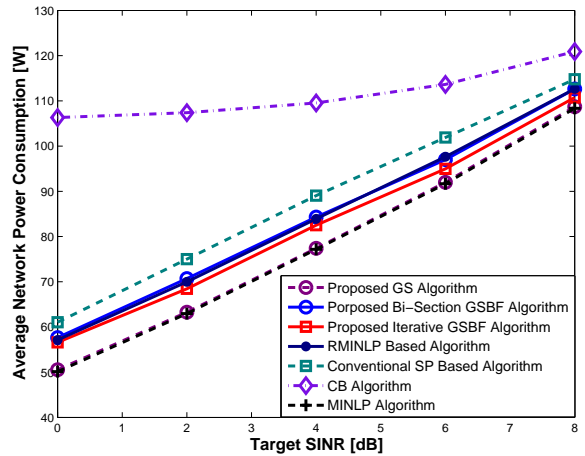}
\caption{Average network power consumption versus target SINR.}
\label{smallnetworktransport}
\end{figure}

Consider a network with $L=10$ 2-antenna RRHs and $K=15$ single-antenna MUs uniformly and independently distributed in the square region $[-1000~1000]\times[-1000~1000]$. We set all the transport link power consumption to be  $P_{l}^{c}=(5+l) W, l=1,\dots,L$, which is to indicate the inhomogeneous power consumption on different transport links.
Fig. {\ref{smallnetworktransport}} demonstrates the average network power consumption with different target SINRs. 

This figure shows that the proposed GS algorithm can always achieve global optimum (i.e., the optimal value from the MINLP algorithm), which confirms the effectiveness of the proposed selection rule for the greedy search procedure. With only logarithmic complexity, the proposed bi-section GSBF algorithm achieves almost the same performance as the RMINLP algorithm, which has a linear complexity. Moreover, with the same complexity, the gap between the conventional SP based algorithm and the proposed bi-section GSBF algorithm is large. Furthermore, the proposed iterative GSBF algorithm always outperforms the RMINLP algorithm,
while both of them have the same computational complexity. These confirm the effectiveness  of the proposed GSBF framework to minimized the network power consumption. Overall, this figure shows that our proposed schemes have the potential to reduce the power consumption by $40\%$ in the low QoS regime, and by $20\%$ in the high QoS regime.

\subsection{Network Power Consumption versus Transport Links Power Consumption}

\begin{figure}[!t]
\centering
\includegraphics[width=1\columnwidth]{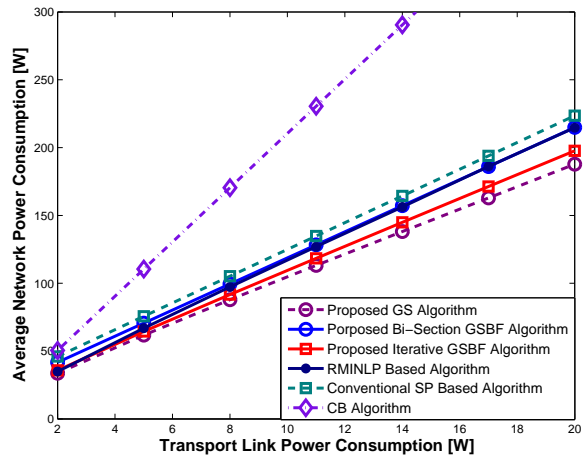}
\caption{Average network power consumption versus transport links power consumption.}
\label{transportlink}
\end{figure}

Consider a network involving $L=20$ 2-antenna RRHs and $K=15$ single-antenna MUs 
uniformly and independently distributed in the square region $[-2000~2000]\times[-2000~2000]$ meters. We assume all the transport links have the same power consumption, i.e., $P_{c}=P_{l}^{c}, \forall l=1,\dots, L$ and set the target SINR as 4 dB. Fig.{\ref{transportlink}} presents average network power consumption with different transport links power consumption.

This figure shows that both the GS algorithm and the iterative GSBF algorithm significantly outperform other algorithms, especially in the high transport link power consumption regime. Moreover, the proposed bi-section GSBF algorithm provides better performance than the conventional SP based algorithm and is close to the RMINLP based algorithm, while with a lower complexity. This result clearly indicates the importance of considering the key system parameters when applying the group sparsity beamforming framework.

\subsection{Network Power Consumption versus the Number of Users} 
\begin{figure}[!t]
\centering
\includegraphics[width=1\columnwidth]{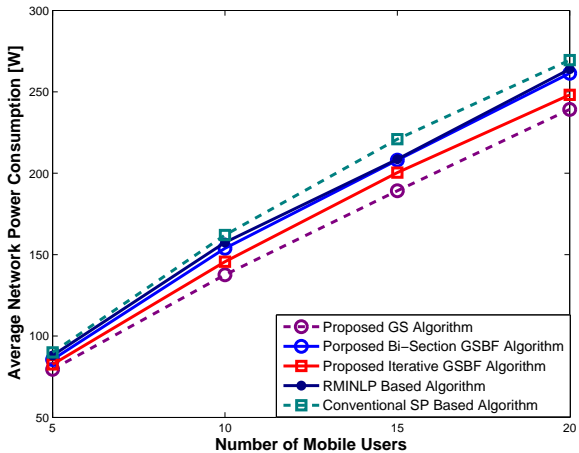}
\caption{Average network power consumption versus the number of mobile users.}
\label{mobileuser}
\end{figure}

Consider a 
network with $L=20$ 2-antenna RRHs uniformly and independently distributed in the square region $[-2000~2000]\times[-2000~2000]$ meters. We assume all
the transport links have the same power consumption, i.e., $P_{l}^{c}=20 W, \forall l=1,\dots, L$ and set the target SINR as 4 dB. Fig. {\ref{mobileuser}} presents the average network power consumption with different numbers of MUs, which are uniformly and independently distributed in the same region.  

Overall, this figure further confirms the following conclusions:
\begin{enumerate}
\item With the $\mathcal{O}(L^{2})$ computational complexity, the proposed GS algorithm has the best performance among all the low-complexity algorithms. \item With the $\mathcal{O}(L)$ computational complexity, the proposed iterative GSBF algorithm outperforms the RMINLP algorithm, which has the same complexity.
\item With $\mathcal{O}(\log(L))$ computational complexity, the proposed bi-section GSBF algorithm has almost the same performance with the RMINLP algorithm and outperforms the conventional SP based algorithm, which has the same complexity. Therefore, the bi-section GSBF algorithm is very attractive for practical implementation in large-scale Cloud-RAN. 
\end{enumerate}


\section{Conclusions and Discussions}
In this paper, we proposed a new framework to improve the energy efficiency of cellular networks with the new architecture of Cloud-RAN. It was shown that the transport network power consumption can not be ignored when designing green Cloud-RAN. By jointly selecting the active RRHs and minimizing the transmit power consumption through coordinated beamforming, the overall network power consumption can be significantly reduced, especially in the low QoS regime. The proposed group sparse formulation $\mathscr{P}_{\textrm{sparse}}$ serves as a powerful design tool for developing low complexity GSBF algorithms. Through rigorous analysis and careful simulations, the proposed GSBF framework was demonstrated to be very effective to provide near-optimal solutions. Especially, for the large-scale Cloud-RAN, the proposed bi-section GSBF algorithm will be a prior option due to its low complexity, while the iterative GSBF algorithm can be applied to provide better performance in a medium-size network. Simulation also showed that the proposed GS algorithm can always achieve nearly optimal performance, which makes it very attractive in the small-size clustered deployment of Cloud-RAN.  

This initial investigation demonstrated the advantage of Cloud-RAN in terms of the network energy efficiency. More works will be needed to exploit the full benefits and overcome the main challenges of Cloud-RAN. Future research directions include more efficient beamforming algorithms for very large scale Cloud-RAN deployment, joint beamforming and compression by considering the limited-capacity transport links, joint user scheduling, and effective CSI acquisition methods.

\appendices
\section{Proof of Proposition 1}
We begin by deriving the tightest positively homogeneous lower bound of $p({\bf{w}})$, which is given by
\cite{Rockafellar1997convex,Obozinski_arXiv2012}
\begin{eqnarray}
p_{h}({\bf{w}})=\inf_{\lambda>0}{{p(\lambda{\bf{w}})}\over{\lambda}}=\inf_{\lambda>0}{{\lambda}}T({\bf{w}})+{1\over{\lambda}}F(\mathcal{T}({\bf{w}})).
\end{eqnarray}
Setting the gradient of the objective function to zero, the minimum is obtained
at $\lambda=\sqrt{F(\mathcal{T}({\bf{w}}))/T({\bf{w}})}$. Thus, the positively
homogeneous lower bound of the objective function becomes
\begin{eqnarray}
\label{lowerbound}
p_{h}({\bf{w}})=2\sqrt{T({\bf{w}})F(\mathcal{T}({\bf{w}}))},
\end{eqnarray}
which combines two terms multiplicatively. 

Define diagonal matrices ${\bf{U}}\in\mathbb{R}^{N\times
N}$, ${\bf{V}}\in\mathbb{R}^{N\times N}$ with $N=K\sum_{l=1}^{L}N_{l}$, and
the $l$-th block elements are $\eta_{l}{\bf{I}}_{KN_{l}}$ and ${1\over{\eta_{l}}}{\bf{I}}_{KN_{l}}$,
respectively.
Next, we calculate the convex envelope of $p_{h}({\bf{w}})$ via computing
its conjugate:
\begin{eqnarray}
p_{h}^{*}({\bf{y}})&=&\sup_{{\bf{w}}\in\mathbb{C}^{N}} \left({\bf{y}}^{T}{\bf{U}}^{T}{\bf{V}}{\bf{w}}-2\sqrt{T({\bf{w}})F(\mathcal{T}({\bf{w}}))}\right),\nonumber\\
&=&\sup_{\mathcal{I}\subseteq\mathcal{V}}\sup_{{\bf{w}}_{\mathcal{I}}\in\mathbb{C}^{|\mathcal{I}|}}
\!\!\!\left({\bf{y}}_{\mathcal{I}}^{T}{\bf{U}}_{\mathcal{I}\mathcal{I}}^{T}{\bf{V}}_{\mathcal{I}\mathcal{I}}{\bf{w}}_{\mathcal{I}}\!-\!2\sqrt{T({\bf{w}}_{\mathcal{I}})F(\mathcal{I})}\right)\nonumber\\
& =& \left\{ \begin{array}{ll}
 0 & {\textrm{if}}~ \Omega^{*}({\bf{y}})\le 1 \\
\infty, & {\textrm{otherwise}}.
  \end{array} \right.
\end{eqnarray}
where ${\bf{y}}_{\mathcal{I}}$ is the $|\mathcal{I}|$-dimensional vector formed with the entries of ${\bf{y}}$ indexed by ${\mathcal{I}}$ (similarly for ${\bf{w}}$), and ${\bf{U}}_{\mathcal{I}\mathcal{I}}$ is the $|\mathcal{I}|\times|\mathcal{I}|$ matrix formed with the rows and columns of ${\bf{U}}$ indexed by $\mathcal{I}$ (similarly for ${\bf{V}}$), and $\Omega^{*}({\bf{y}})$ defines a dual norm of $\Omega({\bf{w}})$:
\begin{eqnarray}
\label{dual1}
\Omega^{*}({\bf{y}})=\sup\limits_{\mathcal{I\subseteq\mathcal{V}}, \mathcal{I}\ne\emptyset}~
{\|{\bf{y}}_{\mathcal{I}}{\bf{U}}_{\mathcal{I}}\|_{\ell_2}\over{2\sqrt{F(\mathcal{I})}}}={1\over{2}}\max_{l=1,\dots,L}
\sqrt{{\eta_{l}}\over{P_{l}^{c}}}{{\|{\bf{y}}_{\mathcal{G}_{l}}\|_{\ell_2}}}.
\end{eqnarray}
The first equality in (\ref{dual1}) can be obtained by the Cauchy-Schwarz inequality:
\begin{eqnarray}
{\bf{y}}_{\mathcal{I}}^{T}{\bf{U}}_{\mathcal{I}\mathcal{I}}^{T}{\bf{V}}_{\mathcal{I}\mathcal{I}}{\bf{w}}_{\mathcal{I}}&\le&\|{\bf{y}}_{\mathcal{I}}{\bf{U}}_{\mathcal{I}}\|_{\ell_2}\cdot\|{\bf{V}}_{\mathcal{I}\mathcal{I}}{\bf{w}}_{\mathcal{I}}\|_{\ell_2}\nonumber\\
&=&\|{\bf{y}}_{\mathcal{I}}{\bf{U}}_{\mathcal{I}}\|_{\ell_2}\cdot\sqrt{T({\bf{w}}_{\mathcal{I}})}.
\end{eqnarray}
The second equality in (\ref{dual1}) can be justified by
\begin{eqnarray}
\Omega^{*}({\bf{y}})&\ge& \sup_{\mathcal{I}\subseteq\mathcal{V},\mathcal{I}\ne\emptyset}\left(
{1\over{2\sqrt{F(\mathcal{I})}}}\max_{l=1,\dots,L}\|{\bf{y}}_{\mathcal{I}\cap\mathcal{G}_{l}}{\bf{U}}_{\mathcal{I}\cap\mathcal{G}_{l}}\|_{\ell_2}\right)\nonumber\\
&=& {1\over{2}}\max_{l=1,\dots,L}\sqrt{{\eta_{l}}\over{P_{l}^{c}}} {{\|{\bf{y}}_{\mathcal{G}_{l}}\|_{\ell_2}}},
\end{eqnarray}
and 
\begin{eqnarray}
\Omega^{*}({\bf{y}})&\le& \sup_{\mathcal{I}\subseteq\mathcal{V},\mathcal{I}\ne\emptyset}\left(
{{\|{\bf{y}}_{\mathcal{I}}{\bf{U}}_{\mathcal{I}}\|_{\ell_2}}\over{2\min_{l=1,\dots,L}\sqrt{F(\mathcal{I}\cap\mathcal{G}_{l})}}}\right)\nonumber\\
&=& {1\over{2}}\max_{l=1,\dots,L}\sqrt{{\eta_{l}}\over{P_{l}^{c}}} {{\|{\bf{y}}_{\mathcal{G}_{l}}\|_{\ell_2}}}.
\end{eqnarray}
Therefore, the tightest convex positively homogeneous lower bound of the
function $p({\bf{w}})$ is 
\begin{eqnarray}
\Omega({\bf{w}})\!\!\!&=&\!\!\!\!\!\!\!\sup_{ \Omega^{*}({\bf{y}})\le1} {\bf{w}}^{T}{\bf{y}}\nonumber\\
\!\!\!\!\!&\le&\!\!\!\!\!\!\!\sup_{\Omega^{*}({\bf{y}})\le1}\sum_{l=1}^{L}\|{\bf{w}}_{\mathcal{G}_{l}}\|_{\ell_2}\|{\bf{y}}_{\mathcal{G}_{l}}\|_{\ell_2}
\nonumber\\
&\le&\!\!\!\!\!\!\!\! \sup_{\Omega^{*}({\bf{y}})\le1} \!\!\left(\!\sum_{l=1}^{L}\sqrt{P_{l}^c\over{\eta_{l}}}\|{\bf{w}}_{\mathcal{G}_{l}}\|_{\ell_2}\right)\!\!\!\left(\!\!\max_{l=1,\dots,L}
\sqrt{{\eta_{l}}\over{P_{l}^{c}}}{{\|{\bf{y}}_{\mathcal{G}_{l}}\|_{\ell_2}}}\right)
  \nonumber\\
&=&\!\!\!\!\! 2\sum_{l=1}^{L}{\sqrt{P_{l}^{c}\over{\eta_{l}}}}\|{\bf{w}}_{\mathcal{G}_{l}}\|_{\ell_2}.
\end{eqnarray}
This upper bound actually holds with equality. Specifically, we let $\bar{\bf{y}}_{\mathcal{G}_{l}}=2\sqrt{P_{l}^c\over{\eta_{l}}}{{\bf{w}}_{\mathcal{G}_{l}}^{\dagger}\over{\|{\bf{w}}_{\mathcal{G}_{l}}^{\dagger}}\|_{\ell_2}}$, such that $\Omega^{*}(\bar{\bf{y}})=1$. Therefore, 
\begin{eqnarray}
\Omega({\bf{w}})&=&\sup_{ \Omega^{*}({\bf{y}})\le1} {\bf{w}}^{T}{\bf{y}}\nonumber\\
&\ge& \sum_{l=1}^{L}{\bf{w}}_{\mathcal{G}_{l}}^{T}\bar{\bf{y}}_{\mathcal{G}_{l}}=2\sum_{l=1}^{L}{\sqrt{P_{l}^{c}\over{\eta_{l}}}}\|{\bf{w}}_{\mathcal{G}_{l}}\|_{\ell_2}.
\end{eqnarray}

\section{Preliminaries on Majorization-Minimization Algorithms}
The majorization-minimization (MM) algorithm, being a powerful tool to find a local optimum by  minimizing a surrogate function that majorizes the objective function iteratively, has been widely used in  statistics, machine learning etc., \cite{hunter2004tutorial}. We introduce the basic idea of MM algorithms, which allows us to derive our main results.

Consider the problem of minimizing $f({\bf{x}})$ over $\mathcal{F}$. The
idea of MM algorithms is as follows. First, we construct a majorization function
$g({\bf{x}}|{\bf{x}}^{[m]})$ for $f({\bf{x}})$ such that  
\begin{eqnarray}
\label{mm1}
g({\bf{x}}|{\bf{x}}^{[m]})\ge f({\bf{x}}), \forall~{\bf{x}}\in\mathcal{F},
\end{eqnarray}  
and the equality is attained when ${\bf{x}}={\bf{x}}^{[m]}$. In a MM algorithm,
we will minimize the majorization function $g({\bf{x}}|{\bf{x}}^{[m]})$ instead
of the original function $f({\bf{x}})$. Let ${\bf{x}}^{[m+1]}$ denote the minimizer of the function $g({\bf{x}}|{\bf{x}}^{[m]})$ over $\mathcal{F}$
at $m$-th iteration, i.e., 
\begin{eqnarray}
\label{mm2}
{\bf{x}}^{[m+1]}=\arg\min_{{\bf{x}}\in\mathcal{F}}~g({\bf{x}}|{\bf{x}}^{[m]}),
\end{eqnarray}
then we can see that this iterative procedure will decrease the value of
$f({\bf{x}})$ monotonically after each iteration, i.e.,
\begin{eqnarray}
f({\bf{x}}^{[m+1]})\le g({\bf{x}}^{[m+1]}|{\bf{x}}^{[m]})\le g({\bf{x}}^{[m]}|{\bf{x}}^{[m]})=f({\bf{x}}^{[m]}),
\end{eqnarray}   
which is a direct result from the definitions (\ref{mm1}) and (\ref{mm2}).
The decreasing property makes an MM algorithm remarkable numerical stability.
 More details can be found in a tutorial on MM algorithms
\cite{hunter2004tutorial} and references therein.



\bibliographystyle{IEEEtran}
\bibliography{/Reference}

\begin{thebibliography}{10}
\providecommand{\url}[1]{#1}
\csname url@samestyle\endcsname
\providecommand{\newblock}{\relax}
\providecommand{\bibinfo}[2]{#2}
\providecommand{\BIBentrySTDinterwordspacing}{\spaceskip=0pt\relax}
\providecommand{\BIBentryALTinterwordstretchfactor}{4}
\providecommand{\BIBentryALTinterwordspacing}{\spaceskip=\fontdimen2\font plus
\BIBentryALTinterwordstretchfactor\fontdimen3\font minus
  \fontdimen4\font\relax}
\providecommand{\BIBforeignlanguage}[2]{{%
\expandafter\ifx\csname l@#1\endcsname\relax
\typeout{** WARNING: IEEEtran.bst: No hyphenation pattern has been}%
\typeout{** loaded for the language `#1'. Using the pattern for}%
\typeout{** the default language instead.}%
\else
\language=\csname l@#1\endcsname
\fi
#2}}
\providecommand{\BIBdecl}{\relax}
\BIBdecl

\bibitem{Soliman_CM13}
I.~Hwang, B.~Song, and S.~Soliman, ``A holistic view on hyper-dense
  heterogeneous and small cell networks,'' \emph{IEEE Commun. Mag.}, vol.~51,
  no.~6, pp. 20--27, Jun. 2013.

\bibitem{Rusek_SPM2013}
F.~Rusek, D.~Persson, B.~K. Lau, E.~Larsson, T.~Marzetta, O.~Edfors, and
  F.~Tufvesson, ``Scaling up {MIMO}: Opportunities and challenges with very
  large arrays,'' \emph{IEEE Signal Process. Mag.}, vol.~30, no.~1, pp. 40--60,
  2013.

\bibitem{Debbah_VTM11}
J.~Hoydis, M.~Kobayashi, and M.~Debbah, ``Green small-cell networks,''
  \emph{IEEE Veh. Technol. Mag.}, vol.~6, no.~1, pp. 37--43, Mar. 2011.

\bibitem{mobile2011c}
{China Mobile}, ``C-{RAN}: the road towards green {RAN},'' \emph{White Paper,
  ver. 2.5}, Oct. 2011.

\bibitem{Wu_WC2012}
J.~Wu, ``Green wireless communications: from concept to reality [industry
  perspectives],'' \emph{IEEE Wireless Commun.}, vol.~19, no.~4, pp. 4--5, Aug.
  2012.

\bibitem{Gesbert_JSAC10}
D.~Gesbert, S.~Hanly, H.~Huang, S.~Shamai~Shitz, O.~Simeone, and W.~Yu,
  ``Multi-cell {MIMO} cooperative networks: A new look at interference,''
  \emph{IEEE J. Sel. Areas Commun.}, vol.~28, no.~9, pp. 1380--1408, Sep. 2010.

\bibitem{WeiYu_WC10}
H.~Dahrouj and W.~Yu, ``Coordinated beamforming for the multicell multi-antenna
  wireless system,'' \emph{IEEE Trans. Wireless Commun.}, vol.~9, no.~5, pp.
  1748--1759, Sep. 2010.

\bibitem{Chang_ICC2013}
C.~Li, J.~Zhang, and K.~Letaief, ``Energy efficiency analysis of small cell
  networks,'' in \emph{Proc. of IEEE Int. Conf. on Commun. (ICC), Budapest,
  Hungary}, Jun. 2013.

\bibitem{Tombaz_GLOBECOM2011}
S.~Tombaz, P.~Monti, K.~Wang, A.~Vastberg, M.~Forzati, and J.~Zander, ``Impact
  of backhauling power consumption on the deployment of heterogeneous mobile
  networks,'' in \emph{Proc. IEEE Global Telecom. Conf. (GLOBECOM)}, Dec. 2011,
  pp. 1--5.

\bibitem{Rao_VT2013}
J.~Rao and A.~Fapojuwo, ``On the tradeoff between spectral efficiency and
  energy efficiency of homogeneous cellular networks with outage constraint,''
  \emph{IEEE Trans. Veh. Technol.}, vol.~62, no.~4, pp. 1801--1814, May 2013.

\bibitem{Donoho_TIT2006}
D.~Donoho, ``Compressed sensing,'' \emph{IEEE Trans. Inf. Theory}, vol.~52,
  no.~4, pp. 1289--1306, 2006.

\bibitem{Tao_IT06}
E.~Candes and T.~Tao, ``Near-optimal signal recovery from random projections:
  Universal encoding strategies?'' \emph{IEEE Trans. Inf. Theory}, vol.~52,
  no.~12, pp. 5406--5425, Dec. 2006.

\bibitem{Berger_CM10}
C.~Berger, Z.~Wang, J.~Huang, and S.~Zhou, ``Application of compressive sensing
  to sparse channel estimation,'' \emph{IEEE Commun. Mag.}, vol.~48, no.~11,
  pp. 164--174, Nov. 2010.

\bibitem{Bach_ML2011}
F.~Bach, R.~Jenatton, J.~Mairal, and G.~Obozinski, ``Optimization with
  sparsity-inducing penalties,'' \emph{Foundations and Trends in Machine
  Learning}, vol.~4, no.~1, pp. 1--106, Jan. 2012.

\bibitem{Ming_SM06}
M.~Yuan and Y.~Lin, ``Model selection and estimation in regression with grouped
  variables,'' \emph{J. R. Statist. Soc. B}, vol.~68, no.~1, pp. 49--67, 2006.

\bibitem{Wainwright_IT2011}
S.~Negahban and M.~Wainwright, ``Simultaneous support recovery in high
  dimensions: Benefits and perils of block
  $\ell_{1}/\ell_{\infty}$-regularization,'' \emph{IEEE Trans. Inf. Theory},
  vol.~57, no.~6, pp. 3841--3863, Jun. 2011.

\bibitem{Z.Q.Luo_JSAC2013}
M.~Hong, R.~Sun, H.~Baligh, and Z.-Q. Luo, ``Joint base station clustering and
  beamformer design for partial coordinated transmission in heterogeneous
  networks,'' \emph{IEEE J. Sel. Areas Commun.}, vol.~31, no.~2, pp. 226--240,
  Feb. 2013.

\bibitem{TonyQ.S._WC2013}
J.~Zhao, T.~Q. Quek, and Z.~Lei, ``Coordinated multipoint transmission with
  limited backhaul data transfer,'' \emph{IEEE Trans. Wireless Commun.},
  vol.~12, no.~6, pp. 2762--2775, Jun. 2013.

\bibitem{Mehanna_SP2013}
O.~Mehanna, N.~Sidiropoulos, and G.~Giannakis, ``Joint multicast beamforming
  and antenna selection,'' \emph{IEEE Trans. Signal Process.}, vol.~61, no.~10,
  pp. 2660--2674, May 2013.

\bibitem{Shamai_TVT2013}
S.-H. Park, O.~Simeone, O.~Sahin, and S.~Shamai, ``Robust and efficient
  distributed compression for cloud radio access networks,'' \emph{IEEE Trans.
  Veh. Technol.}, vol.~62, no.~2, pp. 692--703, 2013.

\bibitem{Shamai_TSP2013}
------, ``Joint precoding and multivariate backhaul compression for the
  downlink of cloud radio access networks,'' \emph{IEEE Trans. Signal
  Process.}, vol.~PP, no.~99, pp. 1--1, 2013.

\bibitem{Jafar_IT2008}
V.~Cadambe and S.~Jafar, ``Interference alignment and degrees of freedom of the
  {$K$}-user interference channel,'' \emph{IEEE Trans. Inf. Theory}, vol.~54,
  no.~8, pp. 3425--3441, Aug. 2008.

\bibitem{Auer_WC2011}
G.~Auer, V.~Giannini, C.~Desset, I.~Godor, P.~Skillermark, M.~Olsson, M.~Imran,
  D.~Sabella, M.~Gonzalez, O.~Blume, and A.~Fehske, ``How much energy is needed
  to run a wireless network?'' \emph{IEEE Wireless Commun.}, vol.~18, no.~5,
  pp. 40--49, Oct. 2011.

\bibitem{Kani_ICP2012}
J.~Kani and H.~Nakamura, ``Recent progress and continuing challenges in optical
  access network technologies,'' in \emph{IEEE 3rd International Conference on
  Photonics (ICP)}, 2012, pp. 66--70.

\bibitem{Dhaini_ITN2013}
A.~Dhaini, P.-H. Ho, G.~Shen, and B.~Shihada, ``Energy efficiency in
  {TDMA}-based next-generation passive optical access networks,''
  \emph{IEEE/ACM Trans. Netw.}, vol.~PP, no.~99, pp. 1--1, 2013.

\bibitem{Shamai_SP2006}
A.~Wiesel, Y.~Eldar, and S.~Shamai, ``Linear precoding via conic optimization
  for fixed {MIMO} receivers,'' \emph{IEEE Trans. Signal Process.}, vol.~54,
  no.~1, pp. 161--176, Jan. 2006.

\bibitem{boyd2004convex}
S.~P. Boyd and L.~Vandenberghe, \emph{Convex optimization}.\hskip 1em plus
  0.5em minus 0.4em\relax Cambridge university press, 2004.

\bibitem{Yuanming_Globecom2013}
Y.~Shi, J.~Zhang, and K.~Letaief, ``Group sparse beamforming for green cloud
  radio access networks,'' in \emph{Proc. IEEE Global Telecom. Conf.
  (GLOBECOM)}, Atlanta, GA, Dec. 2013.

\bibitem{Oppenheim_SP2010}
T.~Baran, D.~Wei, and A.~Oppenheim, ``Linear programming algorithms for sparse
  filter design,'' \emph{IEEE Trans. Signal Process.}, vol.~58, no.~3, pp.
  1605--1617, Mar. 2010.

\bibitem{leyffer_2012mixed}
S.~Leyffer, \emph{Mixed integer nonlinear programming}.\hskip 1em plus 0.5em
  minus 0.4em\relax Springer, 2012, vol. 154.

\bibitem{Cheng_SP2013}
Y.~Cheng, M.~Pesavento, and A.~Philipp, ``Joint network optimization and
  downlink beamforming for {C}o{MP} transmissions using mixed integer conic
  programming,'' \emph{IEEE Trans. Signal Process.}, vol.~61, no.~16, pp.
  3972--3987, 2013.

\bibitem{Obozinski_arXiv2012}
\BIBentryALTinterwordspacing
G.~Obozinski and F.~Bach, ``Convex relaxation for combinatorial penalties,''
  2012. [Online]. Available: \url{http://arxiv.org/abs/1205.1240}
\BIBentrySTDinterwordspacing

\bibitem{Goldsmith_TIT2003}
S.~Vishwanath, N.~Jindal, and A.~Goldsmith, ``Duality, achievable rates, and
  sum-rate capacity of gaussian {MIMO} broadcast channels,'' \emph{IEEE Trans.
  Inf. Theory}, vol.~49, no.~10, pp. 2658--2668, 2003.

\bibitem{hunter2004tutorial}
D.~R. Hunter and K.~Lange, ``A tutorial on {MM} algorithms,'' \emph{The
  American Statistician}, vol.~58, no.~1, pp. 30--37, 2004.

\bibitem{Boyd_2008enhancing}
E.~J. Candes, M.~B. Wakin, and S.~P. Boyd, ``Enhancing sparsity by reweighted
  $\ell_1$ minimization,'' \emph{J. Fourier Anal. Appl.}, vol.~14, no. 5-6, pp.
  877--905, Dec. 2008.

\bibitem{sriperumbudur2011majorization}
B.~K. Sriperumbudur, D.~A. Torres, and G.~R. Lanckriet, ``A
  majorization-minimization approach to the sparse generalized eigenvalue
  problem,'' \emph{Machine learning}, vol.~85, no. 1-2, pp. 3--39, 2011.

\bibitem{Rockafellar1997convex}
R.~T. Rockafellar, \emph{Convex analysis}.\hskip 1em plus 0.5em minus
  0.4em\relax Princeton university press, 1997, vol.~28.

\end{thebibliography}

%
%
%

\end{document}